\documentclass[12pt]{article}
\usepackage{graphics}
\def\one{{\hbox{ 1\kern-.8mm l}}}
\def\ii{{\rm i}}
\def\ex{{\rm e}}
\def\l{\label}

\newcommand{\hepth}[1]{{\tt hep-th/#1}}

\newcommand{\secn}[1]{Section~\ref{#1}}

\newcommand{\ket}[1]{|{#1}\rangle}

\newcommand{\tbl}[1]{Table~\ref{#1}}
\newcommand{\eq}[1]{Eq.~(\ref{#1})}

\def\beq{\begin{equation}}
\def\eeq{\end{equation}}
\def\beqa{\begin{eqnarray}}
\def\eeqa{\end{eqnarray}}
\def\reali{{\hbox{l\kern-.5mm R}}}
\def\compl{{\hbox{l\kern-1.9mm C}}}
\def\interi{{\mathchoice
 {\hbox{Z\kern-1.5mm Z}}
 {\hbox{Z\kern-1.5mm Z}}
 {\hbox{{Z\kern-1.2mm Z}}}
 {\hbox{{Z\kern-1.2mm Z}}}  }}
\newcommand{\sect}[1]{\setcounter{equation}{0}\section{#1}}
\renewcommand{\theequation}{\thesection.\arabic{equation}}
\newcommand{\be}{\begin{equation}}
\newcommand{\ee}{\end{equation}}
\newcommand{\bea}{\begin{eqnarray}}
\newcommand{\ena}{\end{eqnarray}}

\renewcommand{\a}{\alpha}

\newcommand{\shalf}{\frac{1}{2}}

\newcommand{\PRL}[1]{Phys.\ Rev.\ Lett.\ {\bf #1}}

\renewcommand{\thefootnote}{\fnsymbol{footnote}}
\begin{document}
\begin{titlepage}
\rightline{NEIP-9909}
%\rightline{hep-th/9905006}
\rightline{\hfill May 1999}
\vskip 1.2cm
%\vskip 0.8cm
%\centerline{\Large \bf TITLE}
%\vskip 0.5cm
\centerline{\Large \bf Stable non-BPS states in string theory:}
\centerline{\Large \bf a pedagogical review
\footnote{This work is partially supported by the European Commission
TMR programme ERBFMRX-CT96-0045 and by MURST, and is based on the lectures
given at the TMR school {\it Quantum Aspects
of Gauge Theories, Supersymmetry and Unification} in Leuven
(Belgium) January 18-23 1999.}}
\vskip 1.2cm
\centerline{\bf Alberto Lerda $^a$ and Rodolfo Russo $^b$}
\vskip .8cm
\centerline{\sl $^a$ Dipartimento di Scienze e Tecnologie Avanzate}
\centerline{\sl Universit\`a del Piemonte Orientale (Alessandria)}
\centerline{\sl and Dipartimento di Fisica Teorica, Universit\`a di
Torino}
\centerline{\sl and I.N.F.N., Sezione di Torino, Via P. Giuria
1, I-10125 Torino, Italy}
\centerline{\tt lerda@to.infn.it}
\vskip .5cm
\centerline{\sl $^b$ Institut de Phisique, Universit\'e de Neuch\^atel} 
\centerline{\sl Rue A.-L. Breguet 1, CH-2000 
Neuch\^atel, Switzerland}
\centerline{\tt Rodolfo.Russo@iph.unine.ch}
\vskip 1.2cm
\begin{abstract}
We present a pedagogical review of the stable non-BPS states
in string theory which have recently attracted some attention in the
literature. In particular, following the analysis of A. Sen, we discuss in
detail the case of the stable non-BPS D-particle of Type I theory
whose existence is predicted (and required)
by the heterotic/Type I duality. We show that this D-particle originates
from an unstable bound state formed by a D1/anti-D1 pair of Type IIB in which
the tachyon field acquires a solitonic kink configuration.
The mechanism of tachyon condensation is discussed first at a 
qualitative level and then
with an exact conformal field theory analysis.
\end{abstract}
\end{titlepage}
\newpage
\renewcommand{\thefootnote}{\arabic{footnote}}
\setcounter{footnote}{0}
\setcounter{page}{1}
\sect{Introduction}
\label{intro}
\vskip 0.5cm
%%%%%%%%%%%%%%%%%%%%%%%%
% INTRODUCTION
%%%%%%%%%%%%%%%%%%%%%%%%
During the last few years our understanding of string theory
has dramatically increased. Nowadays it is known that all five
consistent superstring models ({\it i.e.} Type IIA, Type IIB,
Type I, $SO(32)$ heterotic and $E_8\times E_8$ heterotic 
\cite{gsw}) and the eleven dimensional supergravity are 
different perturbative expansions of a unique underlying
theory, called M-theory. However, despite 
numerous attempts and remarkable achievements, we are still far from 
a complete and satisfactory
formulation of M-theory. The best we can do at present time
is to define it by means of the non-perturbative duality
relations that connect the various corners of its moduli space 
corresponding to the different string theories.
Establishing (and checking) these relations precisely is therefore 
a necessary step towards a better understanding of M-theory.

One way of testing a duality relation between theory A and theory B
is to show that their effective actions are equivalent up to
a field redefinition. 
In this way, for example, it was discovered that Type IIB
superstring is self-dual \cite{Schwarz}, or that the $SO(32)$ heterotic
string is dual to Type I theory \cite{polcwi}.
Of course, by focusing on the effective
actions, one checks a duality only at a field theory
level, and not in a really stringy way. However, a deeper
understanding can be obtained by looking at the so-called
BPS states. These are a special class of states,
which are characterized
by the key property that their mass is completely determined
by their charge under some gauge field. 
They form short (or ultra-short) multiplets
of the supersymmetry algebra of the theory, and because of
this fact they are {\it stable} and protected from quantum
radiative corrections.
Therefore, their properties can be studied and analyzed
perturbatively at weak coupling in the theory A, and then safely
extrapolated at strong coupling where they can be reinterpreted
in terms of non-perturbative configurations of the dual theory B.
A well-known example of such BPS states is given by
the supersymmetric D$p$ branes of
Type II theories (with $p$ even in Type IIA and $p$ odd 
in Type IIB) which are charged under the $(p+1)$
gauge potentials of the R-R sector \cite{polc95,polc96}.
Since the BPS configurations are protected by non-renormalization
theorems, one may legitimately pose the question
whether the non-perturbative tests based on them are really 
a strong evidence for a duality conjecture or simply unavoidable
results dictated by supersymmetry alone.
Thus, it is of great conceptual importance to analyze a
non-perturbative duality beyond the BPS level.
\par
To this aim, it is useful to observe
that quite often a string theory contains
in its spectrum states which are {\it stable} 
without being BPS. These are in general
the lightest states which carry some conserved quantum numbers
of the theory.
For these states there is no particular 
relation between their mass
and their charge; they receive quantum corrections and
form long multiplets of the supersymmetry algebra
of the theory ({\it i.e.} they are non-BPS).  However, 
being the lightest states with a given set of 
conserved quantum numbers, they are stable since 
they cannot decay into anything else. 
In general it is not difficult to find 
such stable states with the standard perturbative 
methods of string theory and analyze their properties
at weak coupling; but then a very 
interesting question can be asked: 
what happens to these stable
states at strong coupling? Since they cannot decay, 
these states should be present also in the strong coupling
regime, and thus, if the string duality relations 
are correct, one should find stable non-BPS states with 
the right multiplicities and quantum numbers 
also in the dual theory. 
To verify the existence of such states
is therefore a very strong (perhaps the strongest) 
check on a non-perturbative duality relation 
between two string theories.

In these notes, following the analysis of A. Sen
\cite{Sen4,Sen5}, we apply this idea to test the
heterotic/Type I duality \cite{polcwi} at a non-BPS
level \footnote{A first check of this duality has been given in
Ref.~\cite{Dab} where it has been shown that the world-volume theory
on the D-string of Type I is identical to the world-sheet
theory of the heterotic string. Later, further tests involving BPS
saturated terms have been performed in Ref.~\cite{uncomp} for the
uncompactified case and, for the compactified case,
in Ref.~\cite{matal} at level of BPS spectra and in
Ref.~\cite{comp} at level of the effective actions.}.
In this way we get a more complete understanding of this
non-perturbative relation
which, in some sense, has a privileged status: indeed,
a study of the web of string dualities \cite{Sen00}
shows that all of them can be ``derived'' from it
by combining it with T-duality. 

At the first massive level the $SO(32)$ heterotic string contains  
perturbative states which are stable but not BPS.
Their stability follows from the fact that they are the lightest states
carrying the quantum numbers of the spinor representation of the gauge 
group. Since they cannot decay, these states should be 
present also in the strong coupling regime. Then, if the 
heterotic/Type I duality is correct, 
Type I theory should support non-perturbative stable 
configurations that are spinors of SO(32).
It turns out \cite{polB} that a pair formed by a 
D1-brane and an anti D1-brane of Type I 
(wrapped on a circle and with a 
$Z_2$ Wilson line) describes a configuration 
with the quantum numbers of the spinor representation of
SO(32). Thus this system is the right candidate to describe 
in the non-perturbative regime the stable 
non-BPS states of the heterotic string mentioned above.
However, a superposition of a brane with an anti-brane is unstable
due to the presence of tachyons in the open strings
stretching between the brane and the anti-brane \cite{BanksSuss}.
Then the problem of defining properly this superposition and
treating its tachyon instability arises.
This has been addressed by A. Sen in a remarkable series of 
papers \cite{Sen1,Sen2,Sen3,Sen4,Sen5,Sen6}. In particular in 
Ref.~\cite{Sen4} he considered a D-string/anti-D-string
pair of Type IIB, and managed to prove that when the tachyon 
condenses to a kink along the (compact) direction of the
D-string, the pair becomes tightly bound and, as a whole,
behaves as a D-particle. He also computed its mass
finding that it is a factor of $\sqrt{2}$ bigger than the one of the 
supersymmetric BPS D-particle of Type IIA theory. 
The presence of a D-particle in Type IIB spectrum looks
surprising at first sight since one usually thinks that in Type IIB
there are only D$p$ branes with $p$ odd. However, one should keep in mind that
such a D-particle is a non-supersymmetric and non-BPS configuration.
Furthermore it is unstable, due to the fact that there are
tachyons in the spectrum of the open strings living on
its world-line. These tachyons turn out to be odd under the
world-sheet parity $\Omega$, and hence disappear if one performs
the $\Omega$ projection to get the Type I string \cite{TypeI}.
Therefore, the D-particle found by Sen is a stable non-perturbative 
configuration of Type I that transforms as a spinor of SO(32).

Another interesting example of stable non-BPS
configurations that can be used to check a non-perturbative
duality relation is provided by the states that live on the world-volume of a
D5 brane of Type IIB on top of an orientifold $O5$
plane \cite{polB,Sen2,Gab1}. The theory which
supports these states is obtained by
modding out Type IIB theory in ten 
dimensions by $\Omega \,{\cal 
I}_4$ where $\Omega$ is the world-sheet 
parity (the twist operator) and
${\cal I}_4$ is the space-time parity which 
changes the sign of four
of the space-like coordinates, {\it i.e.}
\beq
{\rm IIB} / \Omega \,{\cal I}_4~.
\label{orientifold}
\eeq
This theory contains 32 D5-branes which 
are pairwise located on top of 
16 orientifold 5-planes. 
The open strings that start and end on the two mirror
D5 branes of a given orientifold plane are charged 
under an abelian
gauge potential. It turns out that the lightest 
states which carry 
this abelian charge are at mass-level one, and thus 
are non-BPS but
stable.  To describe these states at strong
coupling, we perform an S-duality transformation 
which maps the twist
parity $\Omega$ to $(-1)^{F_L}$ where $F_L$ 
is the left part of the space-time fermion
number operator. Thus, the S-dual theory of 
(\ref{orientifold}) is the 
orbifold
\beq
{\rm IIB} / (-1)^{F_L} \,{\cal I}_4~,
\label{orbifold}
\eeq
which contains the usual untwisted sector together
with a twisted sector localized on the orbifold fixed plane. 
In particular at the massless level there are, 
among other 
fields, a 2-form Ramond-Ramond potential 
$A^{(2)}$ arising from the 
untwisted R-R sector and a 1-form Ramond-Ramond 
potential $A^{(1)}$
arising from the twisted R-R sector.  
It turns out (see Ref.~\cite{Sen2} for details)
that the consistent
BPS states of this orbifold theory must 
be charged both under $A^{(2)}$ 
and under $A^{(1)}$.
Therefore, they are D1-branes which carry also 
the charge of an abelian vector potential. 
Given this fact, it is
then natural to conjecture that the non-BPS 
states which we are looking
for can be obtained as a superposition of one 
D1-brane and one
anti-D1-brane of this kind. Indeed, by taking 
this combination we cancel
the charge of $A^{(2)}$ and obtain a solitonic
configuration 
which carries only
the charge of an abelian vector potential
and behaves as a D-particle. 
This construction has been performed in Refs.~\cite{Sen2,Gab1}
using the boundary state formalism \cite{clny88}
in which the D-branes are described by a closed
string state (the boundary state) that inserts a boundary
on the string world-sheet and enforces on it the appropriate
boundary conditions~\footnote{For a general discussion of the boundary state
formalism and its application to the study of D-branes and their
interactions see for example Ref.~\cite{bs}.}.
In particular, after treating properly the tachyon instability
on the brane/anti-brane pair in the orbifold model (\ref{orbifold}), 
a stable D-particle state has been explicitly constructed
and its mass has been computed. Of course a similar analysis 
can be done also in other orbifold models \cite{Gab2,Sen1} 
which involve D-branes of different dimensionality
(see for a review Ref.~\cite{Senr}). 

Like in the heterotic/Type I case, also in 
the orbifold theory (\ref{orbifold}) 
one finds that the stable non-BPS states predicted by
duality arise from an unstable system formed by
a D-string/anti-D-string pair of Type IIB in which
the tachyon field acquires a solitonic kink configuration.
Thus, one may wonder whether this mechanism is peculiar to
the two cases we have mentioned, or can occur in more
general situations. This question has been considered
recently by several authors \cite{WITTEN,HORAVAK,Sen6}
using different methods, and new kinds of relation
between BPS and non-BPS branes have been discovered.
It is natural to hope that they may provide new
ideas and clues to understand better the various D-branes and
their fundamental role in string (or M) theory.

For Type II theories these new relations are represented
in Fig. 1. Let us start from a pair formed by a D$p$ brane
and an anti-D$p$ brane of Type IIB ({\it i.e.} $p$ odd) which,
as mentioned above, is an unstable system due to the presence of an
open-string tachyon on its world-volume. If we condense this
tachyon into a solitonic kink solution, then the 
resulting bound state describes
a non-BPS D$(p-1)$ brane of Type IIB \cite{Sen4,WITTEN,Sen6}. 
In these 
notes we will discuss in some detail
only the case $p=1$ which is relevant of the heterotic/Type I duality, but
the same arguments can be used in all other cases as well.
The non-BPS D$(p-1)$ brane of Type IIB constructed in this way
is unstable
because its world volume theory contains again a tachyonic excitation
(which has nothing to do with the previous one). If we repeat
the procedure and make this second tachyon field condense into
a solitonic kink solution, we obtain as a result a D$(p-2)$
brane of Type IIB, which is supersymmetric and stable \cite{Sen6}.
These descent relations, which of course can be established also
with the D-branes of Type IIA theory, are represented  
by the vertical arrows in Fig. 1. 

\begin{figure}[htb]
\begin{center}
%\includegraphicx[scale=.60,angle=270]{d2.eps}
\rotatebox{270}{\scalebox{0.6}{\includegraphics{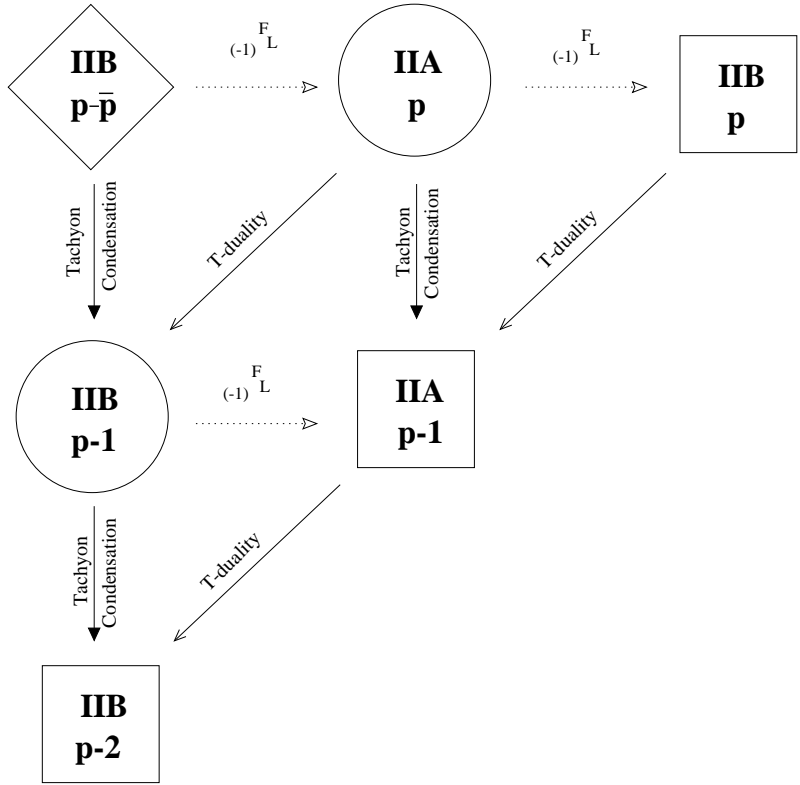}}}
\end{center}
\caption{\baselineskip=13pt {This figure resumes the relations
between different D-branes in type-II superstring theories.
The squares represent the usual supersymmetric BPS
D-branes, while the circles stay for the non-BPS configurations
(unstable in Type II theories).
Starting from a pair formed by a Dp and an anti-Dp brane,
a non-BPS brane can be constructed in two ways: 
one can mod out the system by $(-1)^{F_L}$ (horizontal
arrows) or condense the tachyon living on its 
world-volume (vertical arrows). By repeating these
operations twice, one finds a supersymmetric configuration. 
The diagonal links represent the usual T-duality.}}
\end{figure}

Now let us consider another kind of relations between BPS and non-BPS
branes. Again we start from the D$p$/anti-D$p$ pair of Type IIB and see
what happens when we mod out the system by $(-1)^{F_L}~$
\footnote{The operator $(-1)^{F_L}$ acts on the closed string states
by simply changing the sign of every state that has a left-moving part in the R sector.
Thus, it is an exact symmetry of Type II theories; in particular
$(-1)^{F_L}$ transforms a D$p$ brane into an anti-D$p$ brane and viceversa.
Therefore a brane/anti-brane pair is invariant under this operator
and so it is meaningful to ask what happens when we mod it out.}.
A detailed analysis of the open strings that live on its world-volume
shows that the configuration one obtains after modding out 
$(-1)^{F_L}$ has the same features of
a D$p$ brane of Type IIA. Since $p$ is odd, this must be a non-supersymmetric
and non-BPS brane. By modding out once more by $(-1)^{F_L}$
one goes back to Type IIB theory and finds a stable and 
supersymmetric D$p$ brane \cite{Sen6,Das}. These relations
are represented by the horizontal arrows in Fig.1. 
By exploiting all these relations (as well the
usual T-duality) we can conclude that actually all branes of
Type II theories descend from a bound state of D9 and anti-D9 branes,
a fact that has been used in Refs.~\cite{WITTEN,HORAVAK} to show that the
D-brane charges take values in the K-theory of space time
(see also Refs.~\cite{Garcia,Gukov,Szabo}).

These notes are organized as follows: in \secn{pert}
we briefly summarize the essential perturbative
features of the $SO(32)$ heterotic string and of the Type I theory;
in \secn{d1} we study in some detail the D1 brane of Type I  
and the open string excitations that live on its world-sheet 
and discuss their relations with the heterotic
theory. In \secn{d0} we show how a stable D-particle configuration of 
Type I can be constructed from an unstable D1/anti-D1 pair and study
the mechanism of tachyon condensation from the effective field theory
point of view.
In Sections 5 and 6 we put these results on more solid ground 
by providing an exact conformal description of the
non-BPS D-particle, and finally in Appendix A we
collect a few technical details on the bosonization
procedure which is used in the conformal
field theory analysis. Most of our presentation in the last sections
is based in an essential way on the series of papers by A. Sen
\cite{Sen1,Sen2,Sen3,Sen4,Sen5,Sen6}.

%%%%%%%%%%%%%%%%%%%%%%%%

\vskip 1.5cm
\sect{The SO(32) heterotic string and the Type I theory: a brief 
perturbative description}
\label{pert}
\vskip 0.5cm
%%%%%%%%%%%%%%%%%%%%%%%%%
% insert SECTION 2
%%%%%%%%%%%%%%%%%%%%%%%%%
In this section we are going to summarize the essential
perturbative features of the $SO(32)$ heterotic string and of the Type I theory
that will be relevant for our later
discussions. In subsection 2.1 we briefly analyze
the spectrum of the heterotic string and show
that the states in the spinor representation of $SO(32)$
are non-BPS but stable. Then, in subsection 2.2 we describe
the content of the Type I string viewed as an orientifold of
Type IIB theory. In particular we focus
on the open string sector and also discuss the main
properties of the so-called 1-1 strings that live
on the world-sheet of the D1 branes of the theory.

\vskip 1cm
\subsection{The SO(32) heterotic string} 
\vskip 0.2cm
The heterotic string is a theory of closed oriented strings
with a local gauge invariance in ten dimensions. Roughly 
speaking, it can be
considered as a combination of the bosonic string and the superstring. 
The basic idea of the heterotic construction is to exploit the
holomorphic factorization between the
left and the right moving modes of closed strings and to treat them in
two different ways. In particular, in the right sector one introduces 
the usual field content of the superstring, {\it i.e.} bosonic and 
fermionic coordinates $\widetilde{X}^\mu(\tau\!-\!\sigma)$
and $\widetilde{\psi}^\mu(\tau\!-\!\sigma)$ 
with $\mu=0,1,\ldots,9$ (plus the corresponding ghosts and superghosts),
and performs the standard GSO projection. In the left
sector, instead, one introduces only bosonic coordinates
$X^\rho(\tau\!+\!\sigma)$ with $\rho=0,1,\ldots,25$ 
(plus the corresponding ghosts). 
Since not all $X^\rho$'s can be interpreted as spacetime coordinates, 
the left-moving sector is naturally divided in two sets: when 
$\rho=0,\ldots,9$, the fields
$X^\rho(\tau\!+\!\sigma)$ are combined with 
$\widetilde{X}^\mu(\tau\!-\!\sigma)$ 
to build the usual closed string coordinates $X^\mu(\tau,\sigma)$, while for
$\rho=10,\ldots,25$, the left-moving fields $X^\rho(\tau\!+\!\sigma)$ are 
interpreted as internal degrees of freedom that can propagate on the
world-sheet. 
We will label these sixteen internal coordinates with an index 
$A=1,\ldots,16$.

Given this field content, it is easy to realize that a typical element
of the Fock space of the heterotic theory is a tensor product of a left
and a right state, and looks like
\begin{equation}
\label{state0}
\left(\alpha_{-n_1}^{\mu_1}\cdots\alpha_{-n_\ell}^{A_\ell}\cdots
\ket{k;p}\right)\,\otimes \,\left({\widetilde\alpha}^{\nu_1}_{-m_1}
\cdots{\widetilde\psi}_{-r_p}^{\nu_p}\cdots{\ket{\widetilde{k}}}\right)
\end{equation}
where, in an obvious notation, we have denoted by $\alpha_{-n}$
and ${\widetilde\alpha}_n$
the creation modes of the bosonic fields and by ${\widetilde\psi}_{-r}$
the creation modes of the fermionic coordinates. These ones
have half-integer indices in the NS sector and integer indices in the
R sector. Notice that the left vacuum in \eq{state0} depends on the spacetime
momentum $k^\mu$ as well as on the internal momentum $p^A$, 
whereas the right vacuum
depends only on $k^\mu$. If one considers states in the R sector, the right
vacuum will carry also a spinor label; in these notes however, we will 
not consider explicitly this case and focus only on the NS sector.

The states of the form (\ref{state0}) are not all acceptable 
in the heterotic theory; 
indeed a physical state $\ket{v}$
must satisfy the Virasoro constraints. In particular the mass-shell
condition is
\begin{equation}
\label{hetL0}
\Big(L_0 -1\Big) \ket{v}= \Big(\widetilde{L}_0 - \frac{1}{2}\Big)\ket{v} = 
0~~,
\end{equation}
where we have used the values of the intercept of the bosonic theory 
($a=-1$) in the
left sector and of the NS superstring ($a=-1/2$) in the right sector. 
By expanding
$\widetilde{L}_0$ in modes, one easily finds from the second equality
in \eq{hetL0} 
that the mass $M$ of a state is given by
\begin{equation}
\label{hetmass}
M^2 = \frac{4}{\a'} \Big(\widetilde{N} -\shalf\Big)
\end{equation}
where $\widetilde{N}$ is the total number of right moving
oscillators.
From the first equality of \eq{hetL0} one can derive a generalized
level matching condition which relates 
$\widetilde{N}$ to the total number of left 
moving oscillators $N$ that are present in a given state.
This condition reads as follows
\begin{equation}
\label{lmc}
\widetilde{N} + \shalf  = N + \shalf \sum_A (p^A)^2~,
\end{equation}
where we have conveniently 
measured the internal momenta $p^A$ in units of $\sqrt{2\a'}$.
Further restrictions come from the GSO projection that one has to perform in the
right sector to have a consistent model. In particular, the GSO
projection on NS states selects only 
half-integer occupation numbers $\widetilde N$. Since the
left occupation number $N$ is always integer as 
in the bosonic theory, in order to be able to satisfy \eq{lmc}, 
the internal momenta $p^A$ have to be quantized. 
In particular, the quantity $\sum_A(p^A)^2$ 
has to be an even number. 
This condition implies that the internal coordinates $X^A$ 
must be compactified on an {\it even} 16-dimensional lattice.
The intrinsic consistency of the theory, and more specifically
the modular invariance of the one-loop partition function, requires
that this lattice be also {\it self-dual}. 
It can be shown that there exist only two 16-dimensional
lattices satisfying both these properties: the root lattice of $E_8\times E_8$, 
and a $Z_2$ sublattice of the weight lattice of $SO(32)$. 
Since we are interested in the heterotic theory with gauge group
$SO(32)$, for the rest of these
notes, we will focus only
on the second lattice which is denoted by $\Gamma_{16}$
and is defined by
\begin{equation}
\label{gamma16}
(n_1,\ldots,n_{16})\in \Gamma_{16}~~~\mbox{and}~~~
\Big(n_1+\shalf,\ldots,n_{16}+\shalf\Big)\in \Gamma_{16}
~~\Longleftrightarrow~~
\sum_i n_i \in 2 Z 
\end{equation}
It is interesting to observe that even if in the 
original formulation only a $SO(16)$ symmetry is 
manifest, when the internal coordinates are compactified on 
$\Gamma_{16}$, the resulting gauge group is much bigger. This fact can be
explicitly
checked by counting how many massless states 
involving the internal coordinates $X^A$
are present in the 
theory. Because of \eq{hetmass}, all such states must have ${\widetilde N}=1/2$,
so that their right-moving part is simply 
${\widetilde\psi}_{-1/2}^i{\ket{\widetilde k}}$ where $i=2,...,9$ labels 
the directions transverse to the light-cone.
On the other hand, the level matching condition (\ref{lmc}) requires that 
\begin{equation}
N+\frac{1}{2}\sum_A\left(p^A\right)^2 = 1
\label{lmc1}
\end{equation}
This condition can be satisfied either by taking $N=1$ and $p^A=0$,
or by taking $N=0$ and the momenta $p^A$ to be of the 
form $P= (\pm 1,\pm 1,0,\ldots,0)$ (or any permutation thereof).
With the first choice, we construct 16 states $\alpha_{-1}^A\ket{k;0}$,
while with the second choice we construct 480 states $\ket{k;P}$.
Altogether we have 496 massless states that carry a spacetime vector index
from the right-moving part and span the adjoint representation 
${\bf 496}$ of $SO(32)$.
Thus, they can be legitimately identified with the gauge fields of $SO(32)$. 
The fact that a system of
bosonic fields compactified on an appropriate space exhibits an enlarged
symmetry can be understood 
also from a different point of view through
the bosonization procedure \cite{fms} (see also
Appendix A for some details). In fact by exploiting the 
equivalence
between a compact boson and two fermions, one can substitute the 
original coordinates $X^A$ with 32 fermions $\Lambda^I$ ($I=1,\ldots,32$)
directly in the sigma model 
lagrangian of the heterotic theory, thus making manifest the $SO(32)$ 
symmetry.

It is important to realize that there are 64 more bosonic
massless states in the theory
which correspond to ${\widetilde N}=1/2$,
$N=1$ and $p^A=0$, and are given by
\begin{equation}
\label{state2}
\alpha_{-1}^i\ket{k;0}\,\otimes\,{\widetilde\psi}^j_{-1/2}
{\ket{\widetilde{k}}}
\end{equation}
where the indices $i,j$ run along the transverse directions $2,\cdots,9$.
These states are clearly singlets with respect to the gauge group but are 
space-time tensors. Decomposing them into irreducible components, we 
get a graviton, a dilaton and an antisymmetric two-index tensor. 
In conclusion, the bosonic massless states of the heterotic theory 
are in the following
representations
\begin{equation}
({\bf 1};{\bf 1}) \oplus ({\bf 35};{\bf 1})
\oplus ({\bf 28};{\bf 1}) \oplus ({\bf 8};{\bf 496})
\label{repre1}
\end{equation}
where in each term the two labels refer to the Lorentz 
and gauge groups respectively.
By analyzing the R sector, one finds an equal number of fermionic massless
states that complete the $N=1$ supersymmetric multiplets.

Let us now consider the first excited level of the NS sector 
that consists of states 
with $\widetilde{N}={3}/{2}$ and mass squared $M^2=4/\a'$. There are 128 
ways to realize ${\widetilde N}=3/2$, namely
\begin{eqnarray}
\label{1lr}
&&{\widetilde \psi}^i_{-3/2} 
{\ket{\widetilde{k}}} ~~\to~~\mbox{8 states}\\
\label{2lr}
&&{\widetilde\a}_{-1}^i{\widetilde\psi}^j_{-1/2} 
{\ket{\widetilde{k}}} ~~\to~~\mbox{64 states} \\
\label{3lr}
&& {\widetilde\psi}^i_{-1/2} {\widetilde\psi}^j_{-1/2} 
 {\widetilde\psi}^\ell_{-1/2} {\ket{\widetilde{k}}} ~~\to~~ \mbox{56 states}
\end{eqnarray}
These massive states represent a symmetric two-index tensor 
and an antisymmetric
three-form under the Lorentz group.
In fact, the antisymmetric part of \eq{2lr} together with the states
(\ref{3lr}) completes a massive three-form transforming in the representation
{\bf 84} of the Lorentz group, while the
remaining states transform together as a symmetric two-index tensor
in the representation {\bf 44}. 
The level matching condition (\ref{lmc}) requires to make a tensor product
with left-moving configurations such that
\begin{equation}
N+\frac{1}{2}\sum_A\left(p^A\right)^2 = 2
\label{lmc2}
\end{equation}
There are 73,764 ways to satisfy this requirement! The complete list
of the corresponding states can be found for example on pag. 342 of Ref.
\cite{gsw} where it is shown that they transform as scalars, spinors,
second-rank antisymmetric tensors, fourth-rank antisymmetric tensors and
second-rank symmetric traceless tensors of $SO(32)$. 
Here we focus on the $2^{15}$ states 
that are obtained by taking in \eq{lmc2}
$N=0$ and momenta $p^A$ of the form
$(\pm\shalf,\pm\shalf,\ldots,\pm\shalf)$ with an even
number of plus signs. Notice that these momenta define a point
in the lattice $\Gamma_{16}$, since they satisfy
the second condition of \eq{gamma16}, and correspond
to the {\it spinor} representation of $SO(32)$. Indeed, 
similarly to what happens with the usual spin fields, 
half-integer valued momenta always produce states which are spinors 
of the symmetry group.
By combining these left modes with the 
right-moving ones of Eqs. (\ref{1lr}) - (\ref{3lr}), 
we then obtain bosonic states transforming as
\begin{equation}
\label{trs}
({\bf 44};{\bf 2^{15}}) \oplus ({\bf 84};{\bf 2^{15}})~~.
\end{equation}
Analyzing the first excited level of the R sector, one can find
128 massive fermionic states which transform in the spinor representation
of $SO(32)$ and complete the $N=1$ supersymmetry multiplets.
Thus, altogether the spinors of $SO(32)$ appear with 256 different polarizations,
128 bosonic and 128 fermionic, which correspond to a {\it long}
multiplet of the $N=1$ supersymmetry algebra in ten dimensions. 
Thus, these states cannot satisfy 
the BPS condition \cite{DabHar} but are, nevertheless, stable. Indeed,
as we have seen, at the massless level there are no spinors of $SO(32)$
and thus the states we have found are the lightest ones with these
quantum numbers and hence cannot decay.
As it was observed on pag. 343 of Ref.~\cite{gsw}, this feature makes these
spinors ``absolutely stable, which
might be an interesting prediction for the SO(32) heterotic theory''.
This expectation is going to be confirmed within the non-perturbative
duality between the $SO(32)$ heterotic string and the Type I theory.

\vskip 1cm
\subsection{The Type I string}
\label{typeIs}
\vskip 0.2cm
The Type I string theory is a $N=1$ supersymmetric
model in ten dimensions that consists of both open and closed
{\it unoriented} strings with the usual GSO projection
\footnote{Note that for closed unoriented strings the
GSO projector has to be chiral, like the one of
Type IIB string.}.
It is by now well-known that the Type I theory can be interpreted as 
an orientifold of Type IIB theory, which is a chiral model
consisting only of closed oriented strings
\cite{TypeI}. For a general discussion
on the orientifold projection we refer for example to Ref.~\cite{polB}.
Here it is sufficient to recall that the orientifold projection
is defined by means of a discrete transformation $\Omega$ which in 
general can mix
the world-sheet parity with a space-time operation. Then, a new consistent
spectrum is obtained by removing all states of 
the original theory that are not invariant under $\Omega$, and by 
adding new states that are defined by a generalized
periodicity condition on the string coordinates (the so-called 
twisted sector).

For the Type I theory, the operator $\Omega$ is simply the world-sheet parity, 
namely
\begin{equation}
\Omega~~:~~\sigma~\longrightarrow~\pi-\sigma
\label{parity}
\end{equation}
From this definition, it is easy to derive the action of $\Omega$ on the
string modes; in fact, by expanding the closed string coordinates 
in the usual manner
\begin{equation}
X^{\mu}(\tau,\sigma) = x^\mu +2\alpha'p^\mu\tau
+{\ii}\sqrt{\frac{\alpha'}{2}}
\sum_{n\not= 0}
\left[\frac{\alpha_n^\mu}{n}\,{\rm e}^{-2\ii n(\tau+\sigma)}
+\frac{\widetilde \alpha_n^\mu}{n} \,{\rm e}^{-2\ii n(\tau-\sigma)}
\right]
~~,
\label{Xexpan}
\end{equation}
one immediately sees that $\Omega$ interchanges the left and the right moving
oscillators, {\it i.e.}
\begin{equation}
\Omega \,\alpha_n^\mu\,\Omega^{-1} = \widetilde \alpha_n^\mu
~~~,~~~\Omega \,\widetilde \alpha_n^\mu\,\Omega^{-1} = \alpha_n^\mu
\label{omealpha}
\end{equation}
World-sheet supersymmetry requires that also the fermionic coordinates
$\psi^\mu(\tau+\sigma)$ and $\widetilde\psi^\mu(\tau-\sigma)$ 
transform in a similar way, {\it i.e.}
\begin{equation}
\Omega \,\psi_r^\mu\,\Omega^{-1} = \widetilde \psi_r^\mu
~~~,~~~\Omega \,\widetilde \psi_r^\mu\,\Omega^{-1} = \psi_r^\mu
\label{omepsi}
\end{equation}
where the moding index $r$ is half-integer in the NS sector and
integer in the R sector.
To proceed further, we have to specify also the action of $\Omega$
on the ground states. For simplicity, we choose the $(-1,-1)$ superghost picture 
in the NS-NS sector and the $(-1/2,-1/2)$ superghost picture in the R-R
sector \cite{fms}. Since on the matter fields $\Omega$ simply
exchanges the left and the right sector, it is natural to 
assume that 
\begin{equation}
\Omega\left(\ket{0}_{-1}\otimes \ket{\widetilde 0}_{-1}\right)
= \ket{\widetilde 0}_{-1}\otimes \ket{0}_{-1}
= - \ket{0}_{-1}\otimes \ket{\widetilde 0}_{-1}
\label{omevac}
\end{equation}
for the NS-NS sector, and
\begin{equation}
\Omega\left(\ket{S^\alpha}_{-1/2}\otimes \ket{\widetilde{S}^\beta}_{-1/2}
\right) =  \ket{\widetilde{S}^\alpha}_{-1/2} \otimes
\ket{S^\beta}_{-1/2}
=-\ket{S^\beta}_{-1/2}\otimes \ket{\widetilde{S}^\alpha}_{-1/2}
\label{omevac1}
\end{equation}
for the R-R sector, with $S^\alpha$ and $\widetilde{S}^\beta$
being the left and right spin fields in the chiral spinor representation
of $SO(1,9)$. Notice that the $-$ signs in Eqs. (\ref{omevac}) 
and (\ref{omevac1}) originate from the exchange of the left and right 
vacua 
%$\ket{\widetilde{S}^\alpha}_{-1/2}$ and $\ket{S^\beta}_{-1/2}$
which have fermionic statistics.

Using these rules, it is easy to see which states of Type IIB theory
are even under $\Omega$ and must be kept, and which states are odd and must
be removed. Let us consider for example the massless states of the NS-NS
sector which are represented by
\begin{equation}
\ket{\Phi} = \epsilon_{\mu\nu} \, \psi^\mu_{-1/2}
\,{\widetilde \psi}^\nu_{-1/2}\,\ket{0}_{-1}\otimes
\ket{\widetilde0}_{-1}
 \label{nsns1}
\end{equation}
where the polarization tensor satisfies $k^\mu\epsilon_{\mu\nu}
=\epsilon_{\mu\nu}k^\nu=0$ with $k^2=0$. Then, using 
Eqs. (\ref{omepsi}) and (\ref{omevac}) we
can obtain
\begin{equation}
\Omega\ket{\Phi} = \epsilon_{\nu\mu} \, \psi^\mu_{-1/2}
\,{\widetilde \psi}^\nu_{-1/2}\,\ket{0}_{-1}\otimes \ket{\widetilde 0}_{-1}
\label{nsns12}
\end{equation}
so that only the states with symmetric polarizations are invariant under
$\Omega$. Thus, only the graviton and the dilaton survive the $\Omega$
projection while the antisymmetric Kalb-Ramond field is removed.

Now, let us consider the massless states of the R-R sector which are
represented by 
\begin{equation}
\ket{\Psi} = \frac{1}{(p+2)!} 
\,F_{\mu_0\cdots\mu_{p+2}}
\left(C\Gamma^{\mu_0\cdots\mu_{p+2}}\right)_{\alpha\beta}
\,\ket{S^\alpha}_{-1/2}\otimes \ket{\widetilde{S}^\beta}_{-1/2}
\label{rr1}
\end{equation}
where $C$ is the charge conjugation matrix. This state is physical
if the antisymmetric polarization
tensor satisfies $dF=d^*F=0$ which are precisely the Bianchi identities and the
Maxwell equations for a field strength. Then, using Eqs. (\ref{omepsi}) and
(\ref{omevac1}), we easily find
\begin{equation}
\Omega\ket{\Psi} = -\frac{1}{(p+2)!} 
\,F_{\mu_0\cdots\mu_{p+2}}
\left(C\Gamma^{\mu_0\cdots\mu_{p+2}}\right)_{\beta\alpha}
\,\ket{S^\alpha}_{-1/2}\otimes \ket{\widetilde{S}^\beta}_{-1/2}
\label{rr12}
\end{equation}
so that only those states for which
\begin{equation}
\left(C\Gamma^{\mu_0\cdots\mu_{p+2}}\right)^T = -
\left(C\Gamma^{\mu_0\cdots\mu_{p+2}}\right)
\label{Orr}
\end{equation}
are even under $\Omega$ and must be kept. A little algebra shows that
\eq{Orr} is true if $p=1,5,9$, so that the only R-R potentials that
survive the $\Omega$ projection are a 2-form, its magnetic dual 6-form
and a non-dynamical 10-form. Consequently, in the Type I theory only D1,
D5 and D9-branes can exist as (isolated) brane configurations that carry 
R-R charges.

As we have mentioned before, the closed string 
states which are invariant under $\Omega$
do not represent the whole 
content of the theory. In fact, in the orientifold construction one can
also impose twisted boundary conditions on the string coordinates, that is one 
can require that they be periodic up to an $\Omega$ 
transformation. In our case this amounts to impose that
\begin{equation}
\label{tws}
X^\mu(\tau,\sigma) = \Omega\,X^\mu(\tau,\sigma-\pi)\,\Omega^{-1} = 
X^\mu(\tau,-\sigma) ~~,
\end{equation}
where in the last step we have used the explicit definition
(\ref{parity}). It is easy realize that the twisted boundary 
condition (\ref{tws}) enforces an identification 
between the left and the right-moving parts
since it implies that $\a^\mu_n = \tilde{\a}^\mu_{n}$. Thus, the 
string coordinates which satisfy \eq{tws} and which we denote by
$X^\mu_{\rm N}$, have the same mode expansion
as the coordinates of the {\it open} string with Neumann boundary
conditions, namely
\begin{equation}
\label{opexpan}
X^{\mu}_{\rm N}(\tau,\sigma) = x^{\mu} + (2\alpha^{\prime}) 
p^{\mu}\,\tau + \ii \,\sqrt{2\alpha^{\prime}} \sum_{n \not=
0}{\alpha_{n}^{\mu} \over n} \ex^{-\ii n\tau} \cos\,n\sigma~.
\end{equation}
Then, from \eq{parity} it is immediate to see that the action
of the world-sheet parity on the oscillators of $X_{\rm N}^\mu$
is 
\begin{equation}
\Omega \,\alpha_n^\mu\,\Omega^{-1} = {\rm e}^{\ii n \pi}\,\alpha_n^\mu
\label{OopN1}
\end{equation}
Similarly, one finds that the fermionic oscillators of the open sector 
have the following transformation rule
\begin{equation}
\Omega\,\psi_r^\mu\,\Omega^{-1} = {\rm e}^{\ii r \pi}\,\psi_r^\mu
\label{OopN2}
\end{equation}
To complete our description, we have to specify also the action of $\Omega$
on the ground states of the open sector. For simplicity 
and in analogy with what we have done before, we choose the $(-1)$
superghost picture for the NS sector and the $(-1/2)$ superghost
picture for the R sector
\cite{fms}.
Then, in an obvious notation, we posit that
\begin{equation}
\label{OopN3}
\Omega \ket{0}_{-1} = -\ii\, \ket{0}_{-1}
\end{equation}
and
\begin{equation}
\label{OopN4}
\Omega \ket{S^\alpha}_{-1/2} = - \,\ket{S^\alpha}_{-1/2} 
\end{equation}
With this choice the massless
states of both the NS and R sectors 
are odd under $\Omega$ and thus are removed from the spectrum. This is
a standard result for unoriented open strings {\it without} Chan-Paton factors.
On the contrary, if at the endpoints of the open strings we introduce
Chan-Paton labels for the group $SO(N)$, then the NS and R massless 
states can survive
the $\Omega$ projection. In this case, then, 
the spectrum will contain also $SO(N)$ vector gauge fields 
and their superpartners. However, this theory is consistent only for a particular
choice of the gauge group. In fact, the cancellation of gauge anomalies in the 
low-energy effective action requires that $N=32$ \cite{gs84}. 
This condition can be 
understood also as a massless tadpole cancellation condition in the
Type I string \cite{polcai88}.

This construction can be also formulated in the modern language of 
D-branes and orientifold planes. In this context, one starts again with 
Type IIB theory and, in order to enforce the world-sheet parity
projection, one puts in the background the so-called crosscap state (or
orientifold 9-plane, $O9$) \cite{polcai88,clny87}. This is a 
space-time filling object that is not dynamical, 
since no open strings can end on it, change its shape or switch on a
gauge field. However, the crosscap state is a source of 
closed strings because it can emit, for
example, dilatons, gravitons or a R-R 10-form potential
\footnote{In units where the R-R charge of a D9-brane is $+1$, the R-R 
charge of the $O9$-plane is $-32$.}.
In order to cancel the tadpoles associated to these massless states, one 
is forced to modify the definition of the background of the theory 
by adding 32 D9-branes. This is clearly equivalent to introduce a
sector of open strings with Neumann boundary conditions.
Moreover, since there are 32 different D9 branes on which the open
strings can terminate, it is as if their end-points carry 32 different
charges, thus
inducing the $SO(32)$ gauge group. 

As we have mentioned before, also D1 and D5 branes are compatible
with the orientifold projection (or equivalently with the presence of the
orientifold $O9$-plane). Therefore,
besides the sectors previously considered, 
the Type I theory contains also other sectors corresponding to open strings
defined on these D-branes. Since in particular the open strings living on the
D1 branes of Type I are relevant for
our later discussion, we now discuss their main properties.

Let us consider a supersymmetric D1 brane of Type I, and the open strings 
that live on its two-dimensional world-sheet. These strings, called 
1-1 strings since they have both
end-points on the D1 brane, are characterized by two longitudinal
coordinates $X^a_{\rm N}(\tau,\sigma)$ with Neumann
 boundary conditions ($a=0,1$)
and by eight transverse coordinates $X^i_{\rm D}(\tau,\sigma)$ 
with Dirichlet boundary 
conditions ($i=2,\cdots,9$).
The longitudinal coordinates have the usual mode expansion as in
\eq{opexpan}, while for the transverse coordinates one has
\begin{equation}
\label{opexpanD}
X^{i}_{\rm D}(\tau,\sigma) = y^i + \sqrt{2\alpha^{\prime}} \sum_{n \not=
0}{\alpha_{n}^{\mu} \over n} \ex^{-\ii n\tau} \sin\,n\sigma
\end{equation}
where $y^i$ denotes the position of the D1 brane in the transverse space
\footnote{More generally, if an open string is stretched between two D-branes
located respectively at $y_1^i$ and $y_2^i$, the first term in \eq{opexpanD}
is replaced by $y_1^{i} +\left(\frac{y_2^i-y_1^i}{\pi}\right)
\sigma$.}.
Consequently, the action of $\Omega$ on the ``Dirichlet'' oscillators
is different from the one on the ``Neumann'' oscillators which we have discussed
in the previous subsection. Indeed, using Eqs. (\ref{parity}) and (\ref{opexpanD}),
it is easy to find that
\begin{equation} 
\Omega\,\alpha_n^i\,\Omega^{-1} = - {\rm e}^{\ii n \pi}\,\alpha_n^i
\label{OopD1}
\end{equation}
to be compared with \eq{OopN1}. World-sheet supersymmetry 
requires that the action of $\Omega$ on the fermionic oscillators
in the Dirichlet directions be given by
\begin{equation}
\Omega\,\psi_r^i\,\Omega^{-1} = - {\rm e}^{\ii r \pi}\,\psi_r^i
\label{OopD2}
\end{equation}
for both the NS and R sectors. Finally, on the ground
states of the 1-1 strings one has \cite{polcwi}
\begin{equation}
\Omega \ket{0}_{-1}= -\ii\,\ket{0}_{-1}
\label{OopD3}
\end{equation}
for the NS sector in the -1 superghost picture, and
\begin{equation}
\Omega \ket{S^\alpha}_{-1/2} = - \Gamma^2\cdots\Gamma^9\,
\ket{S^\alpha}_{-1/2}
\label{OopD4}
\end{equation}
for the R sector in the $-1/2$ superghost picture.
\eq{OopD3} is identical to the corresponding one for the
purely Neumann strings since the NS vacuum, being a Lorentz scalar,
is insensitive to the extra minus sign that $\Omega$ picks up
in the Dirichlet directions. On the contrary, the R vacuum is
a (chiral) spinor, and the extra minus sign in the
action of $\Omega$ on the transverse
fermions $\psi^2,\ldots,\psi^9$ is not irrelevant, and indeed it 
translates into the particular structure of 
\eq{OopD4}.

Using these rules, one can easily find that the spectrum of the
1-1 open strings of Type I contains eight massless states both in the
NS and in the R sector. The massless modes of the NS sector
correspond to a vector that accounts for the freedom of translating the
D1 brane along its eight transverse directions. The massless states of the
R sector, instead, correspond to a chiral spinor $\chi$ such that
$\Gamma^2\cdots\Gamma^9\chi=-\chi$, as it follows from \eq{OopD4}. Upon
quantization, the eight components of $\chi$ account for the $2^{8/2}=16$
degeneracy of the D1 brane of Type I, as it should be expected for a 
BPS configuration of the $N=1$ supersymmetry algebra in ten dimensions. 

We conclude by mentioning that in the Type I theory there exist also
open strings with mixed boundary conditions: Neumann at one end-point
and Dirichlet at the other one.  Strings of this kind are for example
the 1-9 open strings that are suspended between a D1 brane and one of
the D9-branes of the Type I background, or the 9-1 strings which have the
opposite orientation. These strings are characterized by eight directions with
mixed boundary conditions, and the states in their spectrum can  
not be eigenvectors of
$\Omega$ since this operator transforms a Neumann-Dirichlet string into a 
Dirichlet-Neumann one. In this case, then,  the orientifold projection
simply requires to keep all symmetric combinations of 1-9 and 9-1 
configurations.
%%%%%%%%%%%%%%%%%%%%%%%%%%
\vskip 1.5cm
\sect{The Type I theory and its D1 brane}
\label{d1}
\vskip 0.5cm
%%%%%%%%%%%%%%%%%%%%%%%%%%
% insert SECTION 3
%%%%%%%%%%%%%%%%%%%%%%%%%%
The $SO(32)$ heterotic string and the Type I theory 
discussed in the previous section look so different 
from each other that they seem completely unrelated. The only evident
similarity between them is that they have the same massless field content. 
However, this fact could be interpreted simply as an accident
due to supersymmetry, because in ten dimensions
the invariance under local supersymmetry completely determines the 
form of the action
for gauge fields coupled to gravity. This interpretation seems to be further
confirmed by the fact that any resemblance 
between the spectra of the two theories disappears
as soon as one goes to the first excited level. For instance, we have mentioned
that in the heterotic theory the states of the first mass level
do not always transform in the same way
under the gauge group, but can be scalars, spinors, second-rank 
symmetric traceless tensors, second-rank or fourth-rank antisymmetric 
tensors of $SO(32)$ \cite{gsw}.
On the contrary, in the Type I theory the color degrees of freedom are
encoded through the Chan-Paton factors (or through the 32 D9-branes of the
Type I background in the modern
language) and thus all perturbative states of this model transform
in the adjoint representation of $SO(32)$. This is a standard 
result that occurs in any string theory where the gauge group is 
introduced by means of the Chan-Paton procedure.

However, despite the appearance, we nowadays know that the $SO(32)$ heterotic
string and the Type I theory describe the same physics \cite{polcwi,wi95},
and are actually the two members of a dual pair of theories related to 
each other in a non-perturbative way.
The first
hint of this equivalence comes from the structure of 
their low-energy effective actions.
Focusing for example on the kinetic 
term of the gauge fields, one has
\begin{equation}
\label{1Hact}
S_{H} \,\sim\,  \int d^{10}x\left[\sqrt{-\det(G_{H})}\,\ex^{-2 \phi_{H}}\, 
{\rm Tr} \,(F_{H}^2) +\ldots
\right]
\end{equation}
for the heterotic string, and
\begin{equation}
\label{2Hact}
S_{I} \,\sim\,  \int d^{10}x\left[\sqrt{-\det(G_{I})} \,\ex^{-\phi_{I}} \,
{\rm Tr} \,(F_{I}^2)+ 
\ldots \right]
\end{equation}
for the Type I theory. In these expressions, $G$ denotes the (string
frame) metric, $\phi$ the dilaton field, and $F$ the $SO(32)$ field strength.
The two actions (\ref{1Hact}) and (\ref{2Hact}) can be mapped into
each other by a field redefinition that involves a change of the metric
\begin{equation}
G^H_{\mu\nu} ~\leftrightarrow~\ex^{-\phi_{I}}G^{I}_{\mu\nu}
\label{metric}
\end{equation}
and a change of the dilaton
\begin{equation}
\phi_{H}~ \leftrightarrow ~ - \phi_{I}
\label{dilaton}
\end{equation}
Since $\ex^{\phi}$ is the string coupling constant, \eq{dilaton}
implies that if there has to be any relation between the heterotic and the
Type I strings, then the strong coupling of one theory has
to be connected to the weak coupling of the other.
This fact also explains why the
two perturbative spectra are so different; indeed, by increasing the
value of the coupling constant, a generic perturbative heterotic state
becomes highly unstable and is not expected to appear in the Type I
theory, and viceversa.

However there are some string configurations that can not decay;
for this reason, they are expected to be present in both theories at
all values of the coupling constant, in particular in the strong 
coupling region of one theory or, equivalently, in the weak 
coupling regime of the dual theory. 
Finding such stable states and checking their multiplicities
is therefore a test of the duality conjecture that
goes well beyond the arguments based on the structure of the effective
actions. A well known class of such stable
configurations consists of the BPS states. 
The main feature of these states is 
that they are annihilated by some supersymmetry generators
so that they carry a charge under one of the gauge fields present in the 
theory and possess a mass which is completely determined by this charge.
Furthermore, these BPS configurations are protected from quantum
radiative corrections, and thus it is particularly simple
to follow their fate when the coupling constant is increased.
For instance, a heterotic string wrapped around a compact dimension 
breaks half the original supersymmetries and is charged under the
antisymmetric two-form present in the gravitational
multiplet,
the charge being simply its winding number.
This is a BPS configuration and hence it should appear in the
Type I theory as a supersymmetric soliton. The natural candidate
for this is the D1 brane of Type I, 
since one expects that the gravitational two-form
gauge field is mapped in the R-R
one, 
just like in the strong/weak duality of Type IIB string. 
Since we are considering BPS configurations, we can immediately test
this proposal. In fact, if it is correct, the tension
\begin{equation}
\tau_H = \frac{1}{2\pi\alpha'}
\label{t1}
\end{equation}
of the fundamental heterotic string should exactly match the tension
\begin{equation}
\tau_{D1} = \frac{1}{2\pi\alpha' g}
\label{t2}
\end{equation}
of the D1 brane of Type I \footnote{These tensions are written using
the normalizations and units of Type II theories.}. 
This is indeed what
happens if we use \eq{metric} to set the same scale in both theories.

The identification between the D1 brane of Type I and the
fundamental string of the heterotic theory can be tested 
at a deeper level; in
fact it is possible to show that the world-sheet structure of a 
(wrapped) heterotic string is exactly reproduced by the world-sheet
dynamics on a (wrapped) D1 brane \cite{polcwi}. Here we recall 
this argument.
In the presence of a D1 brane, the space-time is naturally divided into 
a two-dimensional Minkowski part along the world-sheet of the D1 brane
(spanned by $x^0$ and $x^1$), and an eight-dimensional Euclidean part
transverse to it (spanned by $x^2,\ldots,x^9$). Consequently, the Lorentz
group $SO(1,9)$ is broken into
$SO(1,1)\times SO(8)$. The space transverse to the D1 brane
is going to be identified with the space where the heterotic
string is embedded in the light-cone gauge, while the 
world-sheet of the D1 brane is going to play the role of the world-sheet of the
heterotic string. To see this, we will first consider
the lightest states of the 1-1 open strings attached to the D1 brane
which are responsible for its dynamics, and then identify the corresponding
massless fields with the various string coordinates
of the heterotic theory.
In the NS sector we can construct the following massless states
\begin{equation}
\psi_{-1/2}^\mu\ket{k}_{-1}
\label{NS1}
\end{equation}
with $k^2=0$. Since the momentum $k$ is only along the
longitudinal directions, the field associated to this state will
depend only on $x^0$ and $x^1$. When the index $\mu$ is either
$0$ or $1$ ({\it i.e.} in a NN direction), the state (\ref{NS1})
is odd under $\Omega$ (see Eqs. (\ref{OopN2}) and (\ref{OopN3})), and
thus it is not present in the Type I theory. This means that on the D1 brane
of Type I there are no massless vectors, and thus no local gauge symmetry.
On the contrary, when the index $\mu$ is $2,\ldots,9$ ({\it i.e.}
in a DD direction) the state (\ref{NS1}) is even under $\Omega$
(see Eqs. (\ref{OopD2}) and (\ref{OopD3})), and thus survives
the orientifold projection. The corresponding fields, $\phi^i(x^0,x^1)$
with $i=2,\ldots,9$, are therefore scalars with respect to the world-sheet
group $SO(1,1)$ but are vectors with respect to the transverse group
$SO(8)$. These are the same features of the embedding coordinates
$X^i(\tau,\sigma)$ of the heterotic string, which therefore can be
identified with $\phi^i(x^0,x^1)$ provided that we interpret
$x^0$ and $x^1$ as $\tau$ and $\sigma$ respectively. Notice
since $\sigma$ varies between $0$ and $\pi$, we deduce that the
D1 brane has to be wrapped on a radius $R$ so that $x^1/2R$ can
play the role of $\sigma$.

Let us now consider the massless states in the R sector of the 1-1
strings which are given by the spinor vacua
\begin{equation}
\ket{S^\alpha;k}_{-1/2}
\label{R1}
\end{equation}
with $k^2=0$. The field associated to these states is a chiral spinor
$\chi$ obeying the massless Dirac equation
\begin{equation}
k\!\!\!\slash\chi = \big(k_0\Gamma^0+k_1\Gamma^1\big)\chi=0
\label{dirac1}
\end{equation}
However not all chiral spinors are acceptable in the Type I theory.
In fact, as we have seen in Section 2.2, the orientifold projection
acts on the fermionic zero-modes of the 1-1 strings
as $\Omega=-\Gamma^2\cdots\Gamma^9$, and thus, in order
to be even under $\Omega$, the spinor 
$\chi$ has to satisfy the further constraint
\begin{equation}
\Gamma^2\cdots\Gamma^9 \chi= -\chi
\label{chiral1}
\end{equation}
This condition tells us that in the decomposition of the chiral
spinor representation of $SO(1,9)$ into $(+\frac{1}{2},\bf{8_s})
\oplus(-\frac{1}{2},\bf{8_c})$, only the second term is selected by
the $\Omega$ projection.
Notice that since $\chi$ is chiral, \eq{chiral1} implies that
\begin{equation}
\Gamma^0\Gamma^1\chi = -\chi
\label{chiral2}
\end{equation}
Hence, from the Dirac equation (\ref{dirac1}) we deduce that $k_0=-k_1$,
{\it i.e.} $\chi$ is a right-moving spinor in the $\bf 8_c$
of $SO(8)$. These are precisely the same features of the fermionic coordinates
of the heterotic string in the Green-Schwarz formulation!

In order to complete our description, we still have to
reproduce in the D1 brane
context the color degrees of freedom that in the heterotic theory
are represented by the 32 left-moving fermions resulting from the
fermionization of the 16 internal coordinates.
On the Type I side these degrees of freedom are provided by the 1-9 and 9-1
open strings that stretch between the D1 brane and the 32 
D9-branes of the background. The NS sector of these open strings 
does not contain any massless
excitation, since any physical state in this sector has to satisfy
the Virasoro constraint
\begin{equation}
\Big(L_0 + \frac{1}{2}\Big) \ket{v} = 0~~.
\label{vir1}
\end{equation}
Notice that in this equation we have used the value $a=+1/2$ of the
NS intercept for open strings with eight directions with mixed boundary
conditions \footnote{Remember that the value of the intercept
$a$ can be easily obtained by summing the zero-point energy
of all the light-cone coordinates. For the usual ({\it i.e.} all
NN directions) NS sector one obtains $a=-1/2$, but when there are
$\nu$ directions with mixed boundary conditions one gets $a=-1/2+\nu/8$.}.
Hence in this NS sector even the ground state is massive. 
On the contrary, in the R
sector the world-sheet supersymmetry is preserved 
by the boundary conditions
and the intercept is always vanishing. Therefore, the R ground states are
always massless.
To see which are the other features of these states in the 1-9
and 9-1 sectors, let us recall that the fermions $\psi^0$ and $\psi^1$ 
(which carry an index in a NN direction) have the standard
integral mode expansion and possess zero-modes. Instead, the fermions
$\psi^2,\ldots,\psi^9$ (which carry an index in a direction
with mixed boundary conditions) have a half-integral mode expansion
and do not possess zero-modes. Thus, the R ground states of these
open strings are associated to a field $\Lambda$ 
which is a spinor of $SO(1,1)$ and a scalar of $SO(8)$. Furthermore,
the GSO projection requires that such a spinor be chiral, {\it i.e.}
\begin{equation}
\Gamma^0\Gamma^1 \Lambda = \Lambda
\label{chiral3}
\end{equation}
Combining this condition with the massless Dirac equation 
\begin{equation}
k\!\!\!\slash\Lambda=\big(k_0\Gamma^0+k_1\Gamma^1\big)\Lambda=0
\label{dirac2}
\end{equation}
we deduce that $k_0=k_1$, {\it i.e.} $\Lambda$ is a left-moving
world-sheet spinor! Notice that since $x^1$ is compact,
the momentum $k_1$ in \eq{R1}
must be an integer $n$ in units of $1/R$.

As a matter of fact, this is not enough because we have also to specify
whether we are considering the 1-9 or the 9-1 sector, and distinguish among
the 32 different D9-branes of the background. This information can be
provided by attaching to the state in \eq{R1} appropriate Chan-Paton 
factors, which we denote by $\lambda^I_{19}$ ($I=1,\ldots,32$) for the
1-9 sector, and by $\lambda^I_{91}$ ($I=1,\ldots,32$) for the 9-1 
sector. Since the twist operator $\Omega$ exchanges the string
orientation,
it converts a 1-9 string into a 9-1 string and viceversa, {\it i.e.}
\begin{equation}
\Omega~:~~\lambda_{19}^I ~\leftrightarrow~\lambda_{91}^I
\label{omelam}
\end{equation}
Therefore, the ground state which is selected by the $\Omega$
projection is given by the following symmetric combination
\begin{equation}
\lambda_{19}^I\,\ket{S^+;n}_{-1/2} + \lambda_{91}^I\,\ket{S^+;n}_{-1/2}
\label{symm1}
\end{equation}
By considering all possible values for the momentum (that is by 
summing
over $n$), we can construct a left-moving spinor with an index $I$
in the vector representation of $SO(32)$ and with the
following mode expansion
\begin{equation}
\Lambda^I(\tau,\sigma) = \sum_{n}\Lambda_n^I\,\ex^{\ii 2n(\tau+\sigma)}
\label{lambda1}
\end{equation}
This is precisely the expansion of the 32 heterotic fermions
with {\it periodic} boundary conditions. Upon quantization, these
periodic fermions give rise to the spinorial representations of
$SO(32)$ ({\it  i.e.} those representations with half-integer
$p^A$ in the notation of Section 2.1).
In particular, using only the zero-modes $\Lambda_0^I$, 
we can construct the two representations
${\bf 2^{15}}$ and ${\bf {2^{15}}'}$ of $SO(32)$ that correspond to
$p^A=\pm 1/2$ for all $A$ with an even or odd number
of positive signs respectively. As we have seen in Section 2.1, the 
representation ${\bf 2^{15}}$ appears at the first massive level
of the heterotic string. To obtain the remaining representations
of $SO(32)$, it is necessary to construct 32 fermions with
anti-periodic boundary conditions. To see how this is possible,
we need to understand better the fate of the gauge fields
on the D1 brane of Type I.

We have already seen that the orientation projection forbids to
put a local gauge field on a single D1 brane; however, it is still 
possible that a {\it global} gauge field $A=\frac{\theta}{2\pi R}$ be
compatible with $\Omega$. This constant gauge
field $A$ can always be set to zero,
at least locally, by means of the gauge transformation
\begin{equation}
\label{gt}
A \rightarrow A'=A + \ii \Lambda^{-1} \partial_1\Lambda 
\end{equation}
where $\Lambda=\ex^{\ii {\theta x^1 \over 2\pi R}}~$.
However, even if $A$ is a pure gauge, its presence is not completely 
trivial since it affects the global properties of the charged objects. 
In fact, a field $\Phi$ which carries a charge
$q$ under $A$ and makes a loop around the compact dimension picks up a
non-trivial phase due to the Aharonov-Bohm effect which
is given by
\begin{equation}
\label{wilsonline}
W = \exp{\left\{-\ii q\oint dx^1 A\right\}} = \ex^{-\ii q\theta}~.
\end{equation}
Of course, by gauge invariance, 
this same result has to be obtained also in
the gauge where $A=0$. This is indeed the case because under the gauge 
transformation (\ref{gt}), the charged field $\Phi$ changes as follows
\begin{equation}
\label{gt1}
\Phi \rightarrow \Phi'=\ex^{\ii \frac{q\theta x^1}{2\pi R}}\,
\Phi
\end{equation}
Thus, in the gauge $A=0$, the field of a charged object is no longer
periodic and acquires the non-trivial phase (\ref{wilsonline})
when it is transported around the compact dimension. This fact can 
be rephrased by saying that its momentum $k^1$ is shifted according
to
\begin{equation}
\label{shiftk}
k^1 \rightarrow k^1 + \frac{q\theta}{2\pi R}
\end{equation}

In our specific configuration the field $A$ is localized on the 
D1 brane, and thus any string with at most 
one end-point on the D1 brane is charged under $A$. This is
precisely the case of the 1-9 and 9-1 strings that we are considering. 
By convention we can take the charge to be $q=1$ for the
1-9 strings and to be $q=-1$ for the 9-1 strings.
As we mentioned before, the states in the R sector of these
strings must always appear in linear combinations of the type
\begin{equation}
\label{19-91}
\lambda_{19}^I\, (\ldots)\, \ket{S^+;n}_{-1/2} \,\pm\, 
\lambda_{91}^I\, (\ldots)\, \ket{S^+;m}_{-1/2} ~~,
\end{equation}
where we have explicitly written only the dependence on the momentum
along the compact dimension and understood all other information in the
dots (the sign between the two terms in \eq{19-91} depends on the 
$\Omega$-parity of the oscillator
content). In the absence of any gauge field only those combinations
with $n=m$ are kept. 
However, when there is the constant gauge field 
$A$, things are slightly different. First of all, 
we have to shift the momenta of the charged states
according to \eq{shiftk}, and thus, instead of \eq{19-91},
we must consider the superposition 
\begin{equation}
\label{19-91'}
\lambda_{19}^I\, (\ldots)\, \ket{S^+;n+{\theta}/{2\pi}}_{-1/2} \,\pm
\, \lambda_{91}^I\, (\ldots)\, \ket{S^+;m-{\theta}/{2\pi}}_{-1/2} 
\end{equation}
Since $n$ and $m$ are integers, it is clear that this
combination can be an eigenstate of $\Omega$ only if
$\theta=0$ or $\pi$. Consequently, on the D1 brane of Type I there
exists a $Z_2$ symmetry which is all what remains of the
original $U(1)$ gauge symmetry of the D1 brane of Type IIB
after performing the orientifold projection.
When $\theta=\pi$, the massless fields associated to
\eq{19-91'} are 32 world-sheet spinors $\Lambda^I$
with the following mode expansion
\begin{equation}
\Lambda^I(\tau,\sigma) = \sum_{n}\Lambda_n^I\,\ex^{\ii 
(2n+1)(\tau+\sigma)} 
\label{lambda2}
\end{equation}
This is precisely the expansion of the 32 heterotic fermions with
{\it antiperiodic} boundary conditions. Upon quantization, these
fermions give rise to the integral representations of $SO(32)$
({\it  i.e.} those representations with integer $p^A$ in the
notation of Section 2.1). Representations of this kind are for instance
the singlet ${\bf 1}$, the vector ${\bf 32}$ and the adjoint ${\bf
496}$.

To have the complete equivalence between the low-energy world-sheet
dynamics of the wrapped D1 brane of Type I and the
heterotic string, we still have to impose the appropriate
GSO projection on the 32 heterotic fermions $\Lambda^I$ 
of Eqs. (\ref{lambda1}) and (\ref{lambda2}) in order to remove, for 
example, the representations ${\bf {2^{15}}'}$ and ${\bf 32}$ which 
can be constructed using the $\Lambda^I$'s  but which do not appear
in the heterotic theory at the massless level.   
This reduction can be achieved by using the idea of the orientifold projection
presented in \secn{typeIs}. 
In fact, on the D1 brane of Type I
we can introduce a discrete 
transformation $\Xi$ that acts only on the Chan-Paton factors of the
open strings and is defined as follows: $\Xi$ multiplies by $-1$ each 
time there is an open string ending on the D1 brane and acts trivially otherwise.
Then, it is clear that gauge invariant states ({\it i.e.} 
states that are invariant under the $Z_2$ 
gauge group which survives the $\Omega$ projection) 
are left unchanged by $\Xi$. In view of these considerations,
it appears natural to regard as physical 
excitations of the D-strings only the ones
which are obtained by gauge 
invariant superpositions of 1-9 and 9-1 strings;
in other words, it appears natural to gauge the 
discrete symmetry $\Xi$. We have already encountered an
example of a gauged discrete symmetry in \secn{typeIs}
when we discussed the $\Omega$ projection
and derived unoriented strings from the Type
IIB theory. Here we can apply a similar procedure
to gauge the discrete symmetry $\Xi$.
The first step is to consider only those 
states of the 1-9 and 9-1 spectrum that are invariant under $\Xi$
(the states in the 1-1 and 9-9 sectors are obviously all
invariant under $\Xi$). For example, if we define the spinor representation 
${\bf 2^{15}}$ to be even under $\Xi$, then it is clear that the states
in the other spinor representation ${\bf {2^{15}}'}$
are odd, since they contain one more fermionic zero-mode that
brings a Chan-Paton factor $\lambda_{19}$ or $\lambda_{91}$, and hence 
an extra $-$ sign under $\Xi$.
Thus, the states of ${\bf {2^{15}}'}$ are gauged away and 
the physical degrees of freedom correspond only to 
the representation ${\bf 2^{15}}$, just as the heterotic GSO projection 
requires. However, as we know very well, this is not the whole story, 
since in gauging a discrete symmetry 
one has also to consider
the twisted sector. In
fact, one can impose on the fields living on the
D-string generalized periodicity conditions by exploiting the
transformation $\Xi$, just as we did in \eq{tws} for $\Omega$. In the 
present case, the only fields that can feel $\Xi$ are those in the 1-9 
and 9-1 sectors, {\it i.e.} the $\Lambda^I$'s. Thus,
the twisted sector is characterized by
\begin{equation}
\label{twslambda}
\Lambda^I(\tau,\sigma + \pi) = \Xi\,\Lambda^I(\tau,\sigma) \Xi^{-1} = 
- \Lambda^I(\tau,\sigma)~.
\end{equation}
These are precisely the antiperiodic boundary conditions satisfied by 
the expansion (\ref{lambda2}) which was obtained by
introducing the $Z_2$-Wilson line on the D-string.
Therefore, one can say that introducing this Wilson line is
equivalent to introducing the twisted sector of $\Lambda^I$
under $\Xi$.
Of course, also among the states of the 
twisted sector we have to select  
only the $\Xi$-invariant ones, which 
have an even number of $\Lambda^I$
oscillators, since each one of them contributes with a $-$ sign
to the $\Xi$ parity. Then, one can check that, for example,
the vector representation
${\bf 32}$ is gauged away while the representations ${\bf 1}$ and
${\bf 496}$ are kept, thus reproducing 
the condition $\sum n_i = 2 Z$ of the heterotic side 
(see \eq{gamma16}).
%%%%%%%%%%%%%%%%%%%%%%%%%%
\vskip 1.5cm
\sect{A Non-BPS D-particle in the Type I theory}
\label{d0}
\vskip 0.5cm
%%%%%%%%%%%%%%%%%%%%%%%%%%
% SECTION 4
%%%%%%%%%%%%%%%%%%%%%%%%%%

Up to now, we have focused on BPS objects and exploited their
stability to make a test 
of the non-perturbative duality between the $SO(32)$ heterotic string 
and the Type
I theory. However, string theories often 
contain in their spectrum 
states that are {\it stable} but non-BPS, and thus one may wonder 
whether such states can provide the basis for duality tests that are 
less dependent on supersymmetry.
Let us consider for instance a dual pair of theories $A$ and $B$,
and take all perturbative states of $A$ that 
carry the same set of conserved quantum numbers $\{\alpha\}$.
Clearly, the lightest ones among these states are absolutely stable, 
since any possible decay is energetically
unfavorable, and exist for all values of the
coupling constant of $A$. Hence, if the duality between $A$ and
$B$ is correct, there should exist stable non-perturbative configurations 
of $B$ with the quantum numbers $\{\alpha\}$. Finding such
configurations and checking their multiplicities 
is therefore a significant test of the duality between $A$ and $B$
which does not rely on BPS or supersymmetry arguments. 
In this section we are going to apply this strategy 
and confirm the $SO(32)$ heterotic/Type I
duality by using stable non-BPS objects. Then, in the following section
we will put on more solid ground our results by providing an
exact conformal field theory description of these non-BPS configurations.

We have encountered absolutely stable states when we studied the first 
excited level of the perturbative spectrum of the heterotic string: 
in fact, the massive states in the representation ${\bf 2^{15}}$ of $SO(32)$ 
are the lightest ones transforming as spinors of the gauge group,
since the massless level contains only states in 
the singlet and adjoint representations of $SO(32)$ (see \eq{repre1}). 
Thus, if the heterotic/Type I
duality is correct,
we should find in the 
weakly coupled Type I theory 
some non-perturbative objects  that
behave as massive particles in the spinor representation of $SO(32)$. 
Finding such configurations and providing their explicit
description will be our goals.
As mentioned above, in the heterotic theory
the lightest spinors of $SO(32)$ 
appear at the first massive level and their mass is 
$M=2/\sqrt{\alpha '}$ at $g_H=0$.
As one increases the heterotic coupling constant, the mass $M$ 
gets renormalized, since there are no constraints on it 
coming from supersymmetry, and so, for a general value of $g_{H}$, we 
can write 
\begin{equation}
\label{mass}
M = \frac{2}{\sqrt{\a'}}\,f(g_{H})
\end{equation} 
where the function $f(g_H)$ can in principle be computed in perturbation
theory and is such that $f(0)=1$.
By going to the Type I theory and using the duality relations 
(\ref{metric}) and (\ref{dilaton}), we can 
rewrite the mass $M$ as follows
\begin{equation}
\label{mass1}
M = \frac{2}{\sqrt{\a' g_I}}\,{\widetilde f}(g_{I})
\end{equation} 
where ${\widetilde f} (g)\equiv f(1/g)$.
Thus, contrarily to what happened for the tension of the D-string 
(\ref{t2}), the mass $M$ of these stable non-BPS states
for $g_I\to 0$
is not simply the translation
of the heterotic mass at $g_H=0$, but differs from it by a factor of 
${\widetilde f}(g_{I}\! \to\!0)$.
Of course, such a factor is not accessible to perturbation theory and must be
determined using different methods. 

Let us then examine more carefully
the structure of these non-BPS configurations.
On the heterotic side they are simply excited 
states of the fundamental string; thus, it is natural to expect that 
the D1 brane is involved on the Type I side. However a single 
D1 brane can not reproduce the features of our non-BPS states 
for several reasons. First of all, a D1 brane is a BPS configuration; 
moreover, as we have seen in the previous section, it is dual to a 
fundamental heterotic string with winding number 1. On the contrary, our
non-BPS configurations correspond to {\it unwrapped} strings, {\em i.e.}
to states that are not charged under the gravitational 2-form of the heterotic
theory, and therefore they should be dual to Type I configurations
that are neutral under the R-R 2-form potential. 
The simplest system constructed out of D1 branes but without 
any R-R charge, is clearly the superposition of a D1 brane 
and an anti-D1 brane \footnote{An anti-D$p$ brane is simply a D$p$ brane 
carrying a negative charge under the R-R $p+1$-form potential.}.
However, this configuration is
{\it unstable} due to presence of a tachyonic mode arising from the open
strings stretched between the brane and the anti-brane
\cite{BanksSuss}. To see this, let us study the
brane/anti-brane pair in 
more detail. When we put an anti-D1 brane on top of a D1 brane, we have
four types of  open strings:
\begin{itemize}
\item{the 1-1 strings with both end-points on the D1 brane;}
\item{the ${\bar 1}$-${\bar 1}$ strings with 
both end-points on the anti-D1 brane;} 
\item{the 1-${\bar 1}$ strings starting from the 
D1 brane and ending on the anti-D1 brane;}
\item{the ${\bar 1}$-1 strings starting from the 
anti-D1 brane and ending on
the D1 brane.}
\end{itemize}
As we have seen before, convenient way of encoding 
the information about the end-points of the open strings is by means of
Chan-Paton factors. In the present case we can introduce
$2\times 2$ Chan-Paton matrices $\lambda$ as follows
\begin{eqnarray}
\mbox{for a 1-1 string}&\to& \lambda_{11} = \left(\begin{array}{cc}
 1 & 0 \\ 0 &  0
\end{array}\right)
\label{11} \\
\mbox{for a $\bar 1$-$\bar 1$ string}&\to& 
\lambda_{{\bar 1}{\bar 1}} = \left(\begin{array}{cc}
 0 & 0 \\ 0 &  1
\end{array}\right)
\label{1b1b} \\
\mbox{for a 1-$\bar 1$ string}&\to& \lambda_{1{\bar 1}} = 
\left(\begin{array}{cc}
 0 & 1 \\ 0 &  0
\end{array}\right)
\label{11b} \\
\mbox{for a $\bar 1$-1 string}&\to& \lambda_{{\bar 1}1} 
= \left(\begin{array}{cc}
 0 & 0 \\ 1 &  0
\end{array}\right)
\label{1b1}
\end{eqnarray}
The 1-$\bar 1$ (and $\bar 1$-1) strings differ in
a crucial way from the 1-1 strings considered in Section 2.2 because
they are characterized by a ``wrong'' GSO projection, which, for 
example, selects the tachyonic ground state and removes
the massless vectors in the NS sector. This fact 
can be understood more easily by 
studying this same system from the closed string point of 
view, where the interactions are seen as a tree-level
exchange of closed string states along a cylinder between the brane and 
the anti-brane. However, for the sake of clarity, we first
analyze what happens in the more familiar case of a pair formed by
two D-branes on top of each other. The interaction 
between them is computed by summing the contributions
to the exchange energy of all closed string states 
and can be organized as a sum of the even closed string spin
structures:
NS-NS, NS-NS$(-1)^F$ and R-R ~\footnote{In the cases we are studying 
in this paper, the odd spin structure R-R$(-1)^F$ never gives a 
contribution and thus will not be considered.}.
Of course, this is completely equivalent to
compute the one-loop vacuum energy of open strings 
stretched between the two D-branes, which is 
organized as a sum of the even open string spin structures:
NS, NS$(-1)^F$ and R. In fact, 
by means of a modular transformation $\tau_c\rightarrow \tau_a =
1/\tau_c$, it is possible to map the length $\tau_c$ of the cylinder 
described by the closed strings, into the modular
parameter of the annulus $\tau_a$ spanned by the open strings. 
Then, using the modular properties of the Jacobi $\theta$-functions
that appear in the amplitude, one can explicitly check that under the above
modular transformation the
contribution coming from the exchange of R-R states is exactly mapped
into the contribution of the NS$(-1)^F$ sector of the
1-1 open strings; 
in other words, the exchange of R-R states between two D1 branes
gives the same
result that one finds by computing the one-loop vacuum
energy of the NS sector of the 1-1 open strings 
weighted with the GSO operator $(-1)^F$. The correspondence between
the closed string channel and the open string one can be established
also for the other spin structures, and the results are summarized in 
the following table 
\vskip 0.5cm
\begin{equation}
 \begin{array}{|c|c|}\hline
 \mbox{closed string}     & \mbox{open string}      \\ \hline
 \mbox{NS-NS}      & \mbox{NS}       \\ \hline
 \mbox{NS-NS$(-1)^F$}     & \mbox{R}    \\ \hline
  \mbox{R-R} & \mbox{NS$(-1)^F$}           \\ \hline
 \end{array}
\label{table1}
\end{equation}
\vskip 0.6cm
\noindent
In particular, we easily see that the sum of the NS-NS and R-R sectors 
of the closed string is equivalent to the NS sector of the open string 
with the usual GSO projection. Thus, the NS sector of the
1-1 and $\bar 1$-$\bar 1$ strings contains only states 
that are {\it even} under $(-1)^F$; hence the tachyon ground 
state is not allowed \footnote{As usual, the NS Fock vacuum is taken 
to be odd under $(-1)^F$.}.

Let us now consider an anti-D brane on top of a D brane. In this
case, from the closed string point of view, the NS-NS and the 
NS-NS$(-1)^F$ spin structures give exactly the same contributions
as before, but the contribution of the R-R spin-structure changes sign,
since it is proportional to the product of the R-R charges of the two 
branes. According to table (\ref{table1}), this fact implies that 
the contribution of the NS$(-1)^F$ spin structure of the open
channel changes sign, so that the NS sector of the open strings
stretched between a brane and an anti-brane has the ``wrong'' GSO
projection.
Thus, the NS sector of the 1-${\bar 1}$ and ${\bar 1}$-1 strings 
contains only states that are {\it odd} under $(-1)^F$; in 
particular, the two tachyonic states 
\begin{equation}
\label{ts}
\ket{k}_{-1}\otimes \lambda_{1{\bar 1}}~~~~\mbox{and}~~~~
\ket{k}_{-1}\otimes \lambda_{{\bar 1}1}
\end{equation}
with $k^2=1/2\alpha'$ are permitted.
In conclusion, we can say that the requirement that a cylinder diagram 
can be reinterpreted as an annulus diagram through a modular 
transformation fixes the type of GSO projection that one has to 
perform on the various kinds of open strings stretched between two
branes \footnote{This is also what happens in the so called Type 0 
strings which have recently attracted some attention in the 
literature \cite{type0}.}.
Since these are distinguished by the Chan-Paton factors (see
Eqs. (\ref{11})-(\ref{1b1})), it is natural to introduce a generalized
GSO projection to also take them into account.
Let us then define the operator
$(-1)^{\cal{F}}$ that acts like the usual $(-1)^F$ on the open string 
oscillators and ground states, and that acts on the 
Chan-Paton factors according to
\begin{equation}
\label{gsocp}
(-1)^{\cal{F}}~:~~~~\lambda ~\to~ \sigma^3 \lambda \, \sigma^3~,
\end{equation}
where $\sigma^i$ are the Pauli matrices. Obviously,
$\lambda_{11}$ and $\lambda_{{\bar 1}{\bar 1}}$ are even under  
$(-1)^{\cal{F}}$, while $\lambda_{1{\bar 1}}$ and 
$\lambda_{{\bar 1}1}$ are odd. Thus, our previous
findings can be summarized by saying that the spectrum of the
open strings suspended between two D branes, or between a
D brane and an anti-D brane, contains only states that are
{\it even} under the extended GSO operator  
$(-1)^{\cal{F}}$.

The considerations presented so far apply to any system formed by a 
brane and an anti-brane of Type II, and do not make any use of the
twist operator $\Omega$. Thus, one 
may wonder whether the presence of the tachyons (\ref{ts}) in the 
D1/anti-D1 pair can be avoided by going to the Type I 
theory. We can easily see, however, that this is not the case
because the tachyon instability is not removed by 
the $\Omega$ projection. In fact,
the situation is similar to the one described in \eq{19-91} 
because also here the orientation projection 
selects only a linear combination of the 
two tachyonic states (\ref{ts}). In particular, since the 
oscillator content of the tachyons is even, the combination 
$(\lambda_{1\bar{1}}-\lambda_{\bar{1}1})$ is projected out and
only the symmetric combination of the Chan-Paton matrices 
survives, so that the tachyon in the Type I theory is 
given by
\begin{equation}
\label{tachI}
\ket{T} = \ket{k}_{-1} \otimes (\lambda_{1{\bar 1}} + 
\lambda_{{\bar 1}1}) = 
\ket{k}_{-1} \otimes \sigma^1~~.
\end{equation}

At this point, one has to understand whether the instability due to 
the existence of a tachyonic state in the spectrum is an
incurable problem of the system, or is just a consequence of
the fact that we are
expanding the tachyon field $T$ around a false vacuum. Of course, only 
in this second case one may hope that the D1/anti-D1 pair
could form a bound state that has something to do with the perturbative 
heterotic state we are looking for. 
To give an answer to this question one should compute the
tachyonic potential ${\cal V}(T)$ and see whether or not it
has a minimum. Unfortunately, our knowledge of this potential 
is very limited and thus only a qualitative analysis can be performed
along these lines. Let us nevertheless examine what can be said
in this regard.
First of all, 
from \eq{tachI} 
it is easy to realize that ${\cal V}(T)$ is an even function of $T$.
In fact, let us suppose for a moment that ${\cal V}(T)$ contains a term
like $c\,T^{2n+1}$; then from it one can construct a non-vanishing
amplitude with an odd number of external tachyons which has to be
reproduced by some non-vanishing
string diagram. However, all open string amplitudes
are multiplied by the trace of the Chan-Paton factors of the external
states, which in our case is zero: indeed
$\mbox{Tr}\left[(\sigma^1)^{2n+1}\right]=0$.
Therefore, only even powers
of $T$ are allowed to appear in the tachyonic potential
\footnote{This is to be contrasted with the tachyonic potential of the
bosonic string where also odd powers of $T$ are allowed.}. 
Another thing that is known about ${\cal V}(T)$ is that the coefficient
of the $T^4$ term is positive \cite{BanksSuss,Igor} so that it is conceivable 
to think that there exist a minimum for ${\cal V}(T)$. We now
assume that all higher order terms in the potential do not change
the mexican-hat shape produced by the first two terms, 
and that ${\cal V}(T)$ has an absolute minimum
for $T=\pm T_0$ (since the potential is an even function, the
minima always come in pairs). 

Given these assumptions, we can pose the following question:
if we allow the tachyon to roll down at the values $\pm T_0$ and 
stabilize the system, is it possible that after tachyon condensation
the D1/anti-D1 pair describes the Type I 
D-particle we are looking for? 
The answer to this question is no.
In fact, as we have seen in the previous section,
both the D1 and the anti-D1 branes of Type I carry
the quantum numbers of the spinorial representation of the gauge group,
so that the resulting bound state cannot be a spinor of 
$SO(32)$, which instead is the distinctive feature of the particle
we are searching. On the other hand,
it is quite natural to expect that a brane-antibrane pair
annihilates into 
the vacuum just like a particle-antiparticle pair may do. Actually, this can
happen only if the total energy of the system is vanishing; 
in our particular case
this condition means that 
the negative energy density of the tachyonic condensate has to exactly 
cancel the positive energy density of the two D-strings, {\it i.e.}
\begin{equation}
\label{enecom}
{\cal V}(\pm T_0) + 2 \tau_1 =0~~,
\end{equation}
where $\tau_1$ is the D1 brane tension
given in \eq{t2}. We can rephrase the content of 
\eq{enecom} by saying that when the tachyon
condenses everywhere to $\pm T_0$ the entire system
stabilizes in a ``trivial'' way and reaches a 
configuration that is indistinguishable from the supersymmetric vacuum
into which the D1 and anti-D1 branes have 
annihilated~\footnote{In the literature there are 
other examples in which
tachyon condensation restores space-time supersymmetry;
see for example Ref.~\cite{tachyon}}. 
While this scenario seems quite reasonable at first sight, 
at a deeper analysis it looks too radical and surprising since 
it implies that the only stable configuration in which
the tachyon can condense is the vacuum.
Luckily, as we shall see, there are other possibilities for tachyon condensation
which lead to more interesting results.

The basic problems with the D1/anti-D1 pair we have considered so far
are that both branes are spinors of the gauge group and that the 
resulting system after tachyon condensation 
can have the same quantum numbers of the vacuum. 
However, we know from the previous section 
that there is a simple way to avoid this 
situation. In fact, by compactifying on 
a circle of radius $R$ the direction along which the D-strings
extend and by switching on a $Z_2$ Wilson line, we can remove 
the zero-modes of the 1-9 strings and make the D1 brane 
a scalar of $SO(32)$. 
Hence, a pair formed by a D1 brane and an
anti-D1 brane with a $Z_2$ Wilson line 
({\it e.g.} $\theta=0$ on the D1 brane and $\theta=\pi$
on the anti-D1 brane) has the right quantum numbers to
transform in the spinor representation of $SO(32)$
and is a good candidate to describe the Type I particle.
At first sight, it seems 
unlikely that the presence of a Wilson line 
may change radically the fate of the D1/anti-D1 system; 
in fact, a constant gauge field usually does not modify the 
energy density of a configuration because its field strength is vanishing. 
However, in our case, the situation is quite different:
since our system as a whole transforms like a spinor
of $SO(32)$, the condensation of the tachyonic field can not restore 
the vacuum and a new 
stable configuration with non-zero energy can be realized. 
This is possible because the 
tachyonic state (\ref{tachI}) has only one endpoint on the anti-D1 brane
where $\theta=\pi$ so that
the corresponding field $T$ is charged under the Wilson line and
feels the presence of a non vanishing $\theta$. 
Therefore, the tachyonic condensate is different from the one 
obtained for the system without Wilson line and its contribution to
the energy is not longer that of \eq{enecom}. What is happening
is that the constant gauge field changes the energy density
of the system through the modifications it induces on the charged 
field $T$ which is present in the 
background. To see how this mechanism can work, it is easier to perform
the gauge transformation (\ref{gt}) in order to ``switch off'' the 
Wilson line and deal with a non-interacting system. We have already shown
that in this gauge the wavefunction of an object with charge $q$ is no
longer periodic, but according to \eq{wilsonline}
it acquires a phase ${\rm e}^{-\ii q\theta}$ 
each time it is translated around the wrapped anti-D1 brane. Since in our
case $\theta=\pi$ and $q=1$, 
this simply means that the momentum $k^1$ of the state 
(\ref{tachI}) is quantized in {\it half-integer} multiples of $1/R$ 
so that the tachyonic field $T$ has the following mode expansion
on the D-string world-sheet 
\begin{equation}
\label{tach123}
T(x^0,x^1) = \sum_{n=-\infty}^\infty
T_{n}(x^0)\,{\rm e}^{\,\ii\frac{(n+\frac{1}{2})}{R}\,x^1}
\end{equation}
From this expression we immediately see that 
\begin{equation}
T(x^0,x^1+2\pi R)=-T(x^0,x^1)
\label{tach124}
\end{equation}
{\it i.e.} $T$ satisfies antiperiodic boundary conditions. 
The fact that a bosonic field is not periodic may seem
strange; however this should not come as a surprise
since this is the behavior of any charged field in the presence of a 
Wilson line, while only the neutral ({\em i.e.} gauge invariant) quantities 
remain periodic. Because of \eq{tach124}, a system where $T$ is everywhere 
constant
is no longer acceptable and this shows clearly that the tachyonic condensate 
gets modified by the presence of the $Z_2$ Wilson line.
In fact, what becomes allowed by the new boundary conditions
of $T$
is a configuration that interpolates between the two
minima $\pm T_0$; thus, in this case, the tachyonic condensate
corresponds to a kink solution $T_{\rm kink}$
that passes from $-T_0$ to $T_0$ as one
goes around the compact dimension, namely
\begin{eqnarray}
\label{kink}
T_{\rm kink}(x^1) &\to& -T_0 ~~~~~{\mbox{for}}~~~~~x^1\to -\pi R
\nonumber \\
T_{\rm kink}(x^1) &\to& +T_0 ~~~~~{\mbox{for}}~~~~~x^1\to +\pi R
\end{eqnarray}
Of course, the energy of this configuration 
depends on the exact shape of the tachyonic kink $T_{\rm kink}$
for all values of $x^1$
and, according to our previous arguments, it contributes to the mass 
$M$ of the Type I particle as follows
\begin{equation}
\label{enecom2}
M = \int dx^1 \left[{\cal V}(T_{\rm kink}(x^1)) + 2 \tau_{D1}
\right] 
\end{equation}
Since the kink solution is not explicitly known, we cannot use this equation
to evaluate $M$; however we can easily extract from it the behavior of $M$
for small values of the Type I coupling constant $g_I$. In fact, 
both the classical potential ${\cal V}(T)$ and the brane tension 
$\tau_{D1}$ scale with the usual factor of $1/g_I$
since they originate from a string calculation on a disk. 
Thus the mass $M$ of the non-BPS particle is proportional to the inverse
of the coupling constant, exactly as it
happens for the supersymmetric D-particle 
of Type IIA theory. By using this information in
\eq{mass1}, we can see that for $g_I\to 0$ 
the function ${\widetilde f}(g_I)$ 
is given by
\begin{equation}
{\widetilde f}(g_I)=\frac{c}{\sqrt{g_I}}
\label{f123}
\end{equation}
where $c$ is a constant. In order to fix its value,
we have to leave the field theory analysis and systematically
study the conformal field theory of the open strings living on the 
D1/anti-D1 system. This is what we are going to do in the next section.
However, before concluding this section, we would like to
present one more consideration
at a qualitative level: if one insists that $c$ be a finite constant
for all values of the radius $R$, then $T_{\rm kink}(x^1)$ 
has to be equal to $\pm T_0$
almost everywhere and the core of the kink has
to be concentrated in the $x^1$ 
direction around a point, say $\bar{x}_1$. These
requirements ensure that the energy density in 
\eq{enecom2} is vanishing everywhere because of \eq{enecom}, 
except around the position of the tachyonic soliton. Thus
one can interpret the energy of this system as the mass of a 
{\it localized}
particle. 
Furthermore, the possibility to put the core of the kink in an arbitrary point
$\bar{x}_1$ corresponds to freedom that the particle has to move along the
$x_1$ direction. In the following sections we will show that 
this stable non-BPS particle of Type I is actually a D-particle.

%%%%%%%%%%%%%%%%%%%%%%%%%

\vskip 1.5cm
\sect{Conformal field theory at the critical radius}
\label{cr}
\vskip 0.5cm
%%%%%%%%%%%%%%%%%%%%%%%%%
% insert SECTION 5
%%%%%%%%%%%%%%%%%%%%%%%%%
So far we have not really exploited the presence of the compact 
direction around which the D-strings are wrapped and, in some sense, we
have implicitly taken the $R\to\infty$ limit. In this section, on the
other hand, we consider the D1/anti-D1 system at the particular radius 
\begin{equation}
\label{critrad}
R_c=\sqrt{\a'\over 2}~, 
\end{equation}
and study the conformal field theory of the open strings living
on its world-volume.
In particular, following Ref.~\cite{Sen4}
we show that the non-BPS configuration
corresponding to a D1/anti-D1 bound state 
is a stable D-particle of Type I.
Then, in the next 
section, we will consider the decompactification limit $R\to\infty$
and show that this D-particle is stable
even for $R > R_c$. 
The critical radius $R_c$ is special for several reasons:
first of all, the effective (mass)$^2$ of the scalar state 
(\ref{tachI}), which is tachyonic in the uncompactified space, is 
non-negative. In fact, from the constraint $(L_0 -1/2) \ket{T}=0$, we
find that
\begin{equation}
\label{masslesstach}
M^2_T= \frac{\left(n+{1\over 2}\right)^2}{R^2_c}-\frac{1}{2\a'}\geq 0~,
\end{equation}
where we have taken into account that the momentum along $x_1$ is 
quantized in half-integer units of $R_c$ because of the 
presence of the $Z_2$ Wilson line 
(see also \eq{tach123}).
Moreover, from \eq{masslesstach} it is easy to see that the
lightest excitations of $T$
are those with $n=0$ and $n=-1$ and that, for 
these two values, $M^2_T$ is exactly zero 
\footnote{From these observations we see that 
at the critical radius the field 
$T$ does not represent a true tachyon any more; however,
since it becomes tachyonic in the decompatification
limit, $T$ will be still called tachyon.}. Thus, the conformal 
field theory at $R_c$ possesses two marginal operators
corresponding to these two massless scalar states, which, as
we will see, play a crucial role in our discussion.

A second peculiar feature of the critical radius (\ref{critrad})
is that at $R_c$ the conformal field theory generated by the coordinates
$X^1$ and $\psi^1$ admits several different representations.
In fact, since a bosonic field $X$ 
compactified on a circle of radius $R_c$
and with Neumann boundary conditions is equivalent to a couple
of real NS fermions, the $(X^1$, $\psi^1)$ system is equally well 
described by three fermions $\xi$, $\eta$ and $\psi^1$
where $\xi$ and $\eta$ are defined by
\footnote{In this and in the following formulas we understand
the normal ordering.}
\begin{equation}
\label{bos}
\ex^{\pm{\ii\over \sqrt{2\a'}} X^1} ~~\simeq~ {1\over \sqrt{2}} 
(\xi\pm\ii \eta) ~~~~,~~~~
\eta\,\xi ~~\simeq ~\frac{\ii}{2\alpha'}\partial X^1
\end{equation}
Since the three fermions $\xi$, $\eta$ and $\psi^1$
are on equal footing, we can use again the bosonization 
formulas in order to 
regroup the three fields in a different way and form a new boson.
For example, if we use $\xi$ and $\psi^1$,
we can define a new bosonic field $\phi$
according to
\begin{equation}
\label{bos2} 
\ex^{\pm{\ii\over \sqrt{2\a'}} \phi} ~~\simeq~
{1\over \sqrt{2}} (\xi\pm\ii 
\psi^1)
~~~~,~~~~
\psi^1\,\xi ~~\simeq ~\frac{\ii}{2\alpha'}\partial \phi~~,
\end{equation}
whereas if we use $\eta$ and $\psi^1$, we can introduce yet
another field $\phi'$ by means of 
\begin{equation}
\label{bos3}
\ex^{\pm{\ii\over \sqrt{2\a'}} \phi'} ~~\simeq ~ {1\over \sqrt{2}} 
(\eta\pm\ii \psi^1)
~~~~,~~~~
\psi^1\,\eta ~~\simeq ~\frac{\ii}{2\alpha'}\partial \phi'~~.
\end{equation}
In Appendix A we give some further details of this equivalence
and show that the new fields $\phi$ and $\phi'$ are bosons
compactified on a circle of radius $R_c$; however, 
in most cases, Eqs.~(\ref{bos})-(\ref{bos3}) are sufficient to translate
the physical quantities in the different pictures.
Since all these descriptions are equivalent, one may wonder
what is the advantage of using, for instance, $\phi$ and $\eta$
as independent fields,
instead of the more natural ones $X^1$ and $\psi^1$.  
As we will show momentarily following Ref.~\cite{Sen4}, 
it turns out 
that if one describes the system using 
$\phi$ and $\eta$, it is possible to encode
the tachyonic background explicitly in the conformal
field theory. In order to see why this simplification 
occurs, let us focus on the two massless excitations of the
``tachyonic'' state (\ref{tachI}), {\it i.e.}
\begin{equation}
\label{tach+-} 
\ket{T_\pm} = ~\ex^{\pm{\ii\over \sqrt{2\a'}} X^1} 
\,\ket{0}_{-1}\otimes\sigma_1 \end{equation}
and look at the explicit form of their 
vertex operators ${\cal V}_{T_\pm}$
in the various representations.
Using the bosonization formulas (\ref{bos})-(\ref{bos3})
it is easy to see that 
\begin{equation}
\label{veq1}
{\cal V}_{T_\pm}^{(-1)} \,= \,\ex^{\pm{\ii\over \sqrt{2\a'}} X^1} 
\,\simeq \,
{1\over\sqrt{2}}( \xi \pm \ii \eta )   
\, \simeq \,
\left[\pm{\ii\over\sqrt{2}}\,\eta + 
\frac{1}{2}\left(\ex^{{\ii\over \sqrt{2\a'}}\phi}+\ex^{-{\ii\over 
\sqrt{2\a'}}\phi}\right)\right]
~~,
\end{equation}
where for simplicity we have understood the superghost part
and the Chan-Paton factor $\sigma^1$. From these relations
we immediately realize that the states $\ket{T_\pm}$ of \eq{tach+-}
can also be written either as 
\begin{equation}
\label{seq1}
\ket{T_{\pm}} =
{1\over\sqrt{2}}( \xi_{-{1\over 2}} \pm \ii \,\eta_{-{1\over 2}})
\ket{0}_{-1} \otimes\sigma_1 ~~,
\end{equation}
or as
\begin{equation}
\ket{T_{\pm}} = \left[\pm{\ii\over\sqrt{2}}\,\eta_{-{1\over 
2}}\ket{0}_\phi +{1\over 2}
\left(\ket{+{1\over 2}}_\phi+\ket{-{1\over 2}}_\phi\right)
\right] \ket{0}_{-1}  \otimes\sigma_1~~,
\end{equation}
where we have denoted by $\ket{\ell}_{\phi}$
the vacuum of $\phi$ with momentum $\ell$.
In particular, in the latter representation
the combination 
\begin{equation}
\label{calt}
\ket{\cal T} 
\equiv \frac{1}{\sqrt{2}\,\ii}\Big(\ket{T_+}-\ket{T_-}\Big)= 
\eta_{-{1\over 2}} \ket{0}_\phi \ket{0}_{-1} \otimes\sigma_1
\end{equation}
exhibits a very simple form since it becomes 
formally identical to a
massless vector state at zero momentum in the $-1$
picture with the $\psi$ oscillator 
substituted by $\eta$. This is a first hint to the fact that
the ``tachyonic'' field associated to $\ket{\cal T}$
can be probably treated in the same manner of a gauge
field.

As we have seen in \secn{d1}, 
a constant gauge field in the $X$ direction
can be easily incorporated in a conformal field theory:  
indeed, one can use a {\em free} fields, 
provided that the $X$-momentum of any charged state is 
shifted as indicated in \eq{shiftk}. Therefore, our
working hypothesis is that, at the critical radius $R_c$, 
the ``tachyonic'' background $T_{\rm kink}$
discussed in the previous section can be
described by a Wilson line of the ``gauge'' field 
${\cal T}$ associated to the state
(\ref{calt}). 
In order to identify which is
the bosonic coordinate related to
$\eta$ and write the Wilson line, it is sufficient to look at the vertex operator
corresponding to $\ket{\cal T}$ in the
$0$-superghost picture. 
In the original ($X^1$, $\psi^1$) description, we know 
\cite{fms} that the picture
changing operation transforms the ``tachyon'' 
vertex operators as follows
\begin{equation}
\label{-1/0}
{\cal V}_{T_\pm}^{(-1)}=\ex^{\pm{\ii\over \sqrt{2\a'}}X^1} \to 
{\cal V}_{T_\pm}^{(0)}=\pm \ii\, \psi^1 \ex^{\pm{\ii\over 
\sqrt{2\a'}}X^1}~;
\end{equation}
then, by using the bosonization
formulas (\ref{bos}) and (\ref{bos2}), 
we can easily obtain the $(\phi$, $\eta)$ description of
${\cal V}_{T_\pm}^{(0)}$, and finally get 
\begin{equation}
\label{vat0}
{\cal V}_{\cal T}^{(0)} = {\ii \over \sqrt{2\a'}}\,\partial \phi\,
\otimes\sigma^1~~.
\end{equation}
This is identical to the vertex of the usual gauge boson at zero 
momentum, where $\phi$ plays the role of the coordinate $X$. 
From this equation, it is clear that ${\cal V}_{\cal T}^{(0)}$ represents
a marginal operator which can be used to modify the 
theory. Let us recall that 
at the critical radius $R_c$ our original
D1/anti-D1 pair is not unstable, since there are no 
tachyonic excitations on its world-volume; however, it is always possible 
to deform the conformal field
theory by inserting any number of operators 
${\cal V}^{(0)}_{\cal T}$ without paying any energy. 
This amounts to introduce a Wilson line along
$\phi$ which can be parametrized as follows
\begin{equation}
\label{wilsa}
W(A) = \ex^{{\rm i} {A\over 2\sqrt{2\a'}}\oint d\sigma \, 
\partial_\sigma \phi \,\otimes\,\sigma^1}
\end{equation}
In a qualitative sense, the constant $A$ is equivalent to the tachyon vacuum
expectation value since it multiplies the ``tachyon''
vertex operator ${\cal V}^{(0)}_{\cal T}$.
As we explained in \secn{d1}, the 
states that are charged under a Wilson line 
have a momentum which is shifted according to 
\eq{shiftk} in the gauge where the constant gauge field is zero. 
In our present case, since ${\cal V}^{(0)}_{\cal T}$ is proportional 
to $\sigma^1$, the charged states of the theory are those 
with Chan-Paton factors $\sigma^2$ and 
$\sigma^3$. If we denote their charge by $q$, then
the Wilson line (\ref{wilsa}) induces the
following shift in their $\phi$-momentum $k_\phi$:
\begin{equation}
\label{shift}
k_\phi\;\to\;k_\phi - {Aq\over 2\sqrt{2\a'}} = k_\phi - {q
A\over 4}\,\frac{1}{R_c}
\end{equation}
which implies that the Kaluza-Klein index along $\phi$ is $n-qA/4$
with $n$ integer. 
For later convenience, it is useful to define 
the combinations $\tau^\pm = \sigma^3\pm\ii\sigma^2$ that correspond
to states with charge $q=\mp 2$
under $\sigma^1$:
\begin{equation}
\label{taupm}
[\sigma^1,\tau^\pm] = \mp\, 2\,\tau^\pm~.
\end{equation}
The states with Chan-Paton factors$\one$ and $\sigma^1$ are instead
neutral ($q=0$) under the Wilson line (\ref{wilsa}) and thus
are not affected by the presence of a non-trivial tachyonic background.

For different values of $A$ the conformal field theory on the
world-volume of the D1/anti-D1 pair at the critical radius
describes different physical systems 
which are always characterized by the same mass $M$.
In fact, since at $R_c$ there are no tachyonic modes, the
D1 and the anti-D1 branes with a $Z_2$ Wilson line form
a bound state at threshold whose mass is the integral
over the circle of the energy density of the two
constituents; thus 
\begin{equation}
\label{ormass}
M = 2\,(2\pi R_c)\, \tau_{D1} = 
\frac{\sqrt{2}}{\sqrt{\a'}} {1\over g} = \frac{1}{\sqrt{\a'}\, g_I}~.
\end{equation}
The fact that $M$ is proportional to $1/g_I$ tells us that
this bound state is a D-particle, a result which was
already anticipated in the previous section. Notice that $M$ is a 
factor of $\sqrt{2}$ bigger than the mass of the
supersymmetric (and BPS) D0-brane of Type IIA theory. From
\eq{ormass} we can also fix the value of the constant $c$
defined by \eq{f123} finding $c=1$. This is clearly a result
which is not achievable using perturbative methods!
We remark, however, that it can be obtained using the boundary
state formalism as recently shown in Ref.~\cite{Gallot}.

As long as one studies the system at the critical radius $R_c$, 
all values of $A$ are equally acceptable and lead to a consistent theory;
however in the following section we will show 
that $A=1$ is a preferred value because only for $A=1$
it is meaningful to take the decompactification 
limit $R\to\infty$. Moreover, it turns out that 
at $A=1$ the mass $M$ of the system does not 
depend on the radius \footnote{Notice that this is the same behavior
of the mass that we discussed at the end of the previous section.}.
This is precisely the result that one would 
intuitively expect after tachyon condensation since in that
case the D1/anti-D1 system should really describe a particle. 
However, also in the analysis at the critical radius it is clear that
the case $A=1$ is special and that only for this value of $A$ the system
exhibits the properties required for describing the stable non-BPS
particle we are looking for. 

A first peculiar feature of our system at $A=1$ emerges from
a detailed analysis of the open strings that live on the 
world-volume of the D1/anti-D1 pair. In order to see this, let
us first consider the vacuum energy of the 
open strings in the original system
({\it i.e.} $A=0$) which is defined as
\begin{equation}
\label{vacener}
\Gamma_{A=0} = 2V \int_0^\infty \frac{dt}{2t}
\left[{\rm Tr}_{\rm NS}\left( \frac{1+(-1)^{\cal F}}{2} \,
q^{2L_0-1}\right) - {\rm Tr}_{\rm R}\left( \frac{1+(-1)^{\cal 
F}}{2} \,q^{2L_0}\right)\right]
\end{equation}
where $V$ is the (infinite) length of the time direction 
and $q={\rm e}^{-t}$. The 
traces in \eq{vacener} are to be computed on the NS and R sectors by
taking into account the (extended) GSO projection that acts both
on the string oscillators and on the Chan-Paton factors according to 
\eq{gsocp}. In our present case there is one more thing to take
into account, namely the existence of the $Z_2$ Wilson line on the
anti-D1 brane. This implies that the string states with 
the diagonal Chan-Paton factors$\one$ and $\sigma^3$ (which
have the standard GSO projection on the fermionic oscillators) are
neutral under the $Z_2$ gauge field and their momentum in the
compact direction $X^1$ is quantized in integer units of $1/R_c$.
On the contrary, the states with the off-diagonal Chan-Paton
factors $\sigma^1$ and $\ii\sigma^2$ (which have the
``wrong'' GSO on the fermionic oscillators) carry a $Z_2$
charge and thus their momentum along $X^1$ is quantized in
half-integer units of $1/R_c$. 
This information can be encoded automatically in the formalism
by introducing an operator $h$ which corresponds to a $2\pi R_c$
translation along $X^1$. Its explicit action on the string coordinates
is
\begin{equation}
\label{h12}
h~~:~~~
X^1 ~\to ~ X^1 + 2\pi R_c
~~~,~~~
X^{\mu\not=1} ~\to~ X^{\mu\not=1}
~~~,~~~
\psi^\mu ~\to~ \psi^\mu
\end{equation}
while it acts trivially as the identity on the NS and R ground states.
If we define the action of $h$ on the Chan-Paton factors
$\lambda$ by means of 
\begin{equation}
h~~:~~~\lambda~\to~\sigma^3\,\lambda\,\sigma^3
\label{h123}
\end{equation}
then it is not difficult to realize that only the states which are
even under $h$ belong to the physical spectrum
and contribute to the vacuum energy $\Gamma$. Thus, the presence of the
$Z_2$ Wilson line on the anti-D1 brane can be taken into account by introducing
inside the traces of \eq{vacener} the projector
\begin{equation}
\label{ph}
\Pi_h=\frac{1+h}{2}~~.
\end{equation}
With this in mind, a standard calculation leads to
\begin{eqnarray}
\Gamma_{A=0} & = & 2V\int_0^\infty {dt\over 
2t}\;(8\pi\a't)^{-1/2}\left\{
\sum_{n=-\infty}^{\infty} q^{4n^2}\left[
\left({f_3(q)\over f_1(q)}\right)^8 - \left({f_4(q)\over 
f_1(q)}\right)^8\right] \right. \nonumber \\ \label{pfA0}
&  & +  \sum_{n=-\infty}^{\infty} q^{4 (n+\shalf)^2}
\left[\left({f_3(q)\over f_1(q)}\right)^8 + \left({f_4(q)\over 
f_1(q)}\right)^8\right] \\ \nonumber
& & - \left. \left[\sum_{n=-\infty}^{\infty} q^{4n^2} +
\sum_{n=-\infty}^{\infty} q^{4(n+\shalf)^2}\right]
\left({f_2(q)\over f_1(q)}\right)^8 \right\}~,
\end{eqnarray}
where
\begin{eqnarray}
f_1(q) &=& q^{1 \over 12} \prod_{n=1}^{\infty} (1-q^{2n})
~~~,~~~f_2(q) = \sqrt{2} q^{1 \over 12}
 \prod_{n=1}^{\infty} (1+q^{2n})\nonumber\\
f_3(q) &=& q^{-{1 \over 24}} \prod_{n=1}^{\infty} (1+q^{2n-1})
~~~,~~~
f_4(q)= q^{-{1 \over 24}} \prod_{n=1}^{\infty} (1-q^{2n-1})~~.
\end{eqnarray}
The structure of \eq{pfA0} is quite clear: apart from the 
factor of $(8\pi\a't)^{-1/2}$ coming from the integration over the 
continuum momentum $k_0$ along the time direction, 
the first line is the NS sector contribution
of the open strings with diagonal Chan-Paton factors.
The second line represents instead the contribution coming from the
open strings with off-diagonal Chan-Paton matrices.
Finally, the last line is the contribution of the R sector. 

To perform this same calculation if $A\not=0$, 
it is necessary to use the fields $\phi$ and $\eta$ 
instead of $X^1$ and $\psi^1$, since only in this way we
know how to give the tachyon a non-vanishing vacuum expectation value 
in a conformal field theory framework.
To obtain $\Gamma_A$ it is convenient to start from the $(\phi,\eta)$
description of the system at $A=0$ and select the 
physical states
which are even both under $(-1)^{\cal F}$ and under
$h$. In the $(\phi,\eta)$ representation these discrete symmetries are realized 
explicitly as reported in Tables 1, 2 and 3 of Appendix A.
Then one can switch on the ``tachyonic''
Wilson line and follow how the various states of the spectrum are modified.
In particular, the contributions coming
from the NS open strings with Chan-Paton factors$\one~$ and $\sigma^1$ 
are unchanged, since these two sectors
are neutral under the
Wilson line (\ref{wilsa}) and do not feel the presence of a 
non-trivial ${\cal T}$ background.
Also the contribution of the R sector does not change when $A\not=0$ 
because in this sector
the field $\phi$ obeys mixed Neumann-Dirichlet boundary conditions
so that its momentum is forced always to be zero (see
Appendix A for some details).
Thus, only the NS contribution of the strings with Chan-Paton
factors $\tau^\pm$ 
depends on $A$. Putting all these things together, one finds that
\begin{eqnarray}
\!\!\!& \Gamma_{A} & =~2V \int_0^\infty {dt\over 2t}\;(8\pi\a't)^{-1/2}\left\{
\shalf \sum_{n=-\infty}^{\infty} q^{4n^2}\left[
\left({f_3(q)\over f_1(q)}\right)^8 - \left({f_4(q)\over 
f_1(q)}\right)^8\right] \right. \nonumber \\ \label{pfA1}
\!\!\!&  & + \shalf \sum_{n=-\infty}^{\infty} q^{4(n+\shalf)^2}
\left[ \left({f_3(q)\over f_1(q)}\right)^8 + \left({f_4(q)\over 
f_1(q)}\right)^8\right] \\ \nonumber
\!\!\!& & + \shalf \sum_{n=-\infty}^{\infty} \left[q^{4(n+{A\over 
2})^2} {(f_3(q))^8-(f_4(q))^8\over(f_1(q))^8}
+ \sum_{n=-\infty}^{\infty} 
q^{4(n+\shalf+{A\over 2})^2} 
{(f_3(q))^8+(f_4(q))^8\over(f_1(q))^8}\right] \\ 
\nonumber
\!\!\!& & - \left. \left[\sum_{n=-\infty}^{\infty} q^{4n^2} +
\sum_{n=-\infty}^{\infty} q^{4(n+\shalf)^2}\right]
\left({f_2(q)\over f_1(q)}\right)^8 \right\}~~.
\end{eqnarray}
The $A$ dependence is only in
the third line which represents the contribution of the states with
Chan-Paton factors $\tau^\pm$ whose charge under $W(A)$ is 
$\mp 2$ (see \eq{taupm}). In particular,
the first term with an exponent proportional to $(n+{A\over 2})^2$, comes 
from the states with an odd number of $\eta$-oscillators, while the second
term with an exponent proportional to $(n+\shalf+{A\over 2})^2$, is 
produced by the states with an even number of $\eta$-oscillators.

At $A=1$ a remarkable simplification occurs: all terms proportional to 
$f_4(q)$ cancel. This means that, for this value of $A$ there is no
NS$(-1)^F$ spin structure in the open string channel and consequently,
according to \eq{table1}, there is no R-R contribution to
the vacuum energy from the closed string point of view \footnote{The cancellation
of the R-R sector can also be seen directly in the closed string
description by using the boundary state approach \cite{Gallot}.}! 
Thus, only for $A=1$ the D1/anti-D1 pair really 
represents a neutral object under the R-R field and can be identified with
the perturbative heterotic states with zero winding number.  

Another important property of the system at $A=1$ is that it becomes
so tightly bound that it is not possible to separate
the two
constituent D1-branes without paying energy. 
To see this, let us consider the massless NS states
\begin{equation}
\label{scst}
\psi^i_{-{1\over 2}} \ket{0}_{-1}~~~
\hbox{with}~~i=2,\ldots,9~~,
\end{equation}
which describe the shape of a D-string, and see 
the modification induced on them by the tachyon background. 
Notice that actually we can
construct two different states out of \eq{scst} by
tensoring it with the Chan-Paton factors$\one~$ or $\sigma^3$ 
(the other two Chan-Paton matrices $\sigma^1$ and $\ii\sigma^2$
are removed by the GSO projection).
The first combination is neutral under the Wilson line
$W(A)$
and thus remains massless for all values of $A$; 
it corresponds to the marginal operators
\begin{equation}
\label{maral}
{\cal V}^{i,(-1)} = \psi^i\,\otimes\one
\end{equation}
for $i=2,...,9$ which account for the
freedom of rigidly moving the D1/anti-D1 pair
in the flat transverse directions $X^i$. 
On the other hand, the operator
$\psi^i_{-{1\over 2}} \otimes\sigma^3$ 
represents a translation of the two branes 
in opposite directions and thus is related to the possibility of 
separating the two components of the system. However, being 
proportional to $\sigma^3$, this type of state feels the effect of the
tachyon background: if one introduces the matrices $\tau^\pm$, 
it is easy to find the modification induced by $A$, namely
\begin{eqnarray}
\label{scstp}
\psi^i_{-{1\over 2}} \ket{0}_{-1}\,\ket{0}_\phi\otimes
\sigma^3 &=& \frac{1}{2} \psi^i_{-{1\over 2}} \ket{0}_{-1}\,\ket{0}_\phi\otimes
(\tau^+ + \tau^-)
\\ 
&\rightarrow& \shalf\,
\psi^i_{-{1\over 2}} \ket{0}_{-1}  \left(
\ket{-{A\over 2}}_\phi\otimes \tau^+ + 
\ket{\,{A\over 2}}_\phi\otimes \tau^- \right)~.
\nonumber
\end{eqnarray}
When $A\not = 0$ this state acquires a mass proportional to
$A^2$ thus showing that there are no zero-modes permitting
the separation of the D-string and the anti-D-string. 
Note that the r.h.s. of \eq{scstp} is {\em not} invariant under the
discrete transformation $h$ corresponding to a $2\pi R_c$
translation along the compact direction. However, this is not surprising
because we know from \secn{d0} that the periodicity of charged
states is modified by the presence of a Wilson line: in our case,
from Eqs.~(\ref{shift}) and (\ref{wilsonline}), we can see that
a state of charge $q$ picks up a phase equal to $\ex^{-\ii {\pi A q\over 2}}$
when it is transported once around the circle.
Since in our normalizations all charged states
have $|q| =2$, we see that
also the value $A=2$ is particular: indeed, at $A=2$ all 
states of the spectrum (even the charged ones!) have the same periodicity of the
original theory at $A=0$. Moreover, at $A=2$ 
the vacuum energy (\ref{pfA1}) reduces to that of \eq{pfA0} confirming the fact
that the conformal field theories at $A=0$ and at $A=2$ describe 
the same physical situation. 

This analysis shows that, at the critical radius $R_c$, the presence of the
``tachyonic'' background can be nicely encoded in the conformal
field theory by introducing
a constant vacuum expectation value 
$A$ for the field ${\cal T}$
(see Eqs.~(\ref{calt}) and (\ref{vat0})).
Moreover,
$A$ turns out
to be an angular variable ranging from $0$ to $2$. The middle point of
this interval ($A=1$) has two special properties: 
\begin{itemize}
\item at $A=1$ the system as a whole does not emit any R-R field, so 
that it is possible to identify it with a heterotic state not 
charged under $B_{\mu\nu}$
at strong coupling;
\item at $A=1$ the D-string and the anti-D-string are most
tightly bound, since the lightest state describing the separation of the 
two has maximum mass. In fact, when $A>1$, one can consider a state
similar to that of \eq{scstp}, but proportional to $\left(
\ket{1-{A\over 2}}_\phi\otimes \tau^+ + 
\ket{-1+ {A\over 2}}_\phi\otimes \tau^- \right)$; it has the right
periodicity to be present in the physical spectrum and is lighter than
the one of \eq{scstp} with $A=1$.
\end{itemize}
These observations suggest that the value $A=1$  really represents
the minimum of the tachyonic potential which corresponds to the 
kink solution discussed in the previous section. Thus, one expects
that this value of $A$ should give rise to a stable 
system also when we take the decompactification limit $R\to\infty$ and
recover the full $SO(9)$ invariance. We note however, 
that even at the critical
radius $R_c$ 
it is possible to find a ninth zero mode corresponding to the
freedom of moving the ${\cal T}$ soliton along the $x^1$ direction. 
Under a translation $x^1\to x^1+\epsilon$, the
vertex operators (\ref{veq1}) transform as
\begin{equation}
\delta_\epsilon {\cal V}_{T_\pm}^{(-1)}
= \pm \ii\, \epsilon\;
\ex^{\pm{\ii\over \sqrt{2\a'}} X^1} \otimes\sigma^1~.
\end{equation}
so that
\begin{eqnarray}
\label{v123}
\delta_\epsilon {\cal V}_{\cal T}^{(-1)} &=&
\frac{1}{2\ii}\left(\delta_\epsilon {\cal V}_{T_+}^{(-1)}
-\delta_\epsilon {\cal V}_{T_-}^{(-1)}\right)
\\
&=&\frac{\epsilon}{\sqrt{2}}\left(\ex^{+{\ii\over \sqrt{2\a'}} X^1}+
\ex^{-{\ii\over \sqrt{2\a'}} X^1}
\right)\otimes\sigma^1 \simeq \epsilon\,\xi\,\otimes\sigma^1
\nonumber
\end{eqnarray}
Thus, the marginal vertex operator
\beq \l{9zm}
{\cal V}^{(-1)} = \xi  
\otimes{\sigma^1}
\eeq
represents the ninth zero mode we were looking for and
can be combined with the eight vertices (\ref{maral})
to generate the $SO(9)$ symmetry associated to the
freedom of moving the D-particle in space. 

So far the twist operator $\Omega$ has not 
played any significant role.
However, we conclude this section by stressing that 
the $\Omega$ projection is actually a
crucial ingredient in the entire construction.
In fact, even if it is possible to study the D1/anti-D1 pair in the context of
Type IIB strings, one does not expect to find in this case a {\it
stable} configuration corresponding to a D-particle. This is
because Type IIB theory is self-dual under S-duality and 
there is no place for a stable non-BPS D-particle 
configuration. On the other hand, 
the instability of the D1/anti-D1 pair
(with a constant gauge field $\theta=1$) in the Type
IIB theory is clearly displayed also at the level of the conformal field
theory.
In fact, if the worldsheet parity projection is not taken into 
account, one has to consider all open
string states of the 1-1, ${\bar 1}$-${\bar 1}$,
1-${\bar 1}$ and ${\bar 1}$-1 sectors which are even
under $(-1)^{\cal F}$ and h, without requiring that they
be even under $\Omega$.
In our particular case, this implies
that the ``tachyonic'' field ${\cal T}$ becomes complex, because
besides the state in \eq{tachI}, there is also the combination with
a Chan-Paton factor $\ii\sigma^2$. 
Moreover, the usual $U(1)\times U(1)$ gauge
bosons
of any brane/anti-brane pair with Chan-Paton factors$\one$ and $\sigma^3$
are also present in the spectrum
of the oriented theory. 
At the critical radius $R_c$ and at $A=0$, these four massless states 
can be written in terms of $\eta$ and
$\phi$ according to
\begin{eqnarray}
\label{4ns}
&&\ex^{\pm{\ii\over \sqrt{2\a'}}X^1}\ket{0}_{-1}\otimes
\ii\sigma^2 
~
\to ~ \left[\pm{\ii\over\sqrt{2}}\,\eta_{-{1\over 2}}\ket{0}_\phi +{1\over 2}
\left(\ket{\,\shalf}_\phi+\ket{-{1\over 2}}_\phi\right)
\right] \ket{0}_{-1}  \otimes\ii\sigma_2~, \nonumber
\\
&&\psi_{-\frac{1}{2}}^1\ket{0}_{-1}\otimes\sigma^3 ~\to~
\left[-{\ii\over \sqrt{2}} 
\left(\ket{\,\shalf}_\phi-\ket{-{1\over 2}}_\phi
\right) \right] \ket{0}_{-1}  \otimes\sigma_3~, \\
&&\psi_{-\frac{1}{2}}^1\ket{0}_{-1}\otimes\one  ~\to ~
\left[-{\ii\over \sqrt{2}} 
\left(\ket{\,\shalf}_\phi-\ket{-{1\over 2}}_\phi
\right) \right] \ket{0}_{-1}  \otimes \one~. 
\nonumber
\end{eqnarray}
We remark that these states are odd under $\Omega$
and thus do exist in the Type I theory~\footnote{It is worth pointing out that the action of the
twist operator $\Omega$ on the NS vacuum of the 1-${\bar 1}$
and ${\bar 1}$-1 sectors is given by
\[
\Omega\,\ket{0}_{-1} = \ket{0}_{-1}
\]
{\it i.e.} there is an extra factor of $\ii$ with
respect to \eq{OopN3} which refers to 
the 1-1 sector. This fact can be understood using
for example the arguments presented in Ref.~\cite{PolGim};
see also Ref.~\cite{Gallot}.}.
By taking a linear combination of the first two lines in \eq{4ns} 
we obtain the following state
\beq\l{ntac}
\ket{\tt t}_{A=0}  = \left(\ket{\,\shalf}_\phi \otimes\tau^+  - \ket{-\shalf}_\phi
\otimes\tau^-\right) \ket{0}_{-1}~.
\eeq
The presence of the Chan-Paton factors $\tau^\pm$ shows 
that this state
is charged under the tachyonic Wilson line $W(A)$. 
The other possible combinations that can be constructed
out of \eq{4ns} are instead neutral. 
In particular, if we set $A=1$ which is the value corresponding to 
the tachyonic kink solution, the shift (\ref{shift})
of the $\phi$-momentum
transforms \eq{ntac} into a very simple expression
\beq\l{ntac2}
\ket{\tt t}_{A=1} = \ii\,\ket{0}_\phi \ket{0}_{-1} \otimes\sigma^2~.
\eeq
It is not difficult to realize that this state has
a {\it negative} mass$^2$, {\it i.e.} it is tachyonic!
Thus, even if we have condensed the original tachyon field
${\cal T}$, the Type IIB system responds by creating a new 
tachyonic state. Thus, in Type IIB theory,
there is an unavoidable 
tachyonic instability also at the critical radius whenever one 
gives a non-vanishing vacuum expectation value to the field ${\cal T}$. This should
not be surprising because, as we have said before, the 
tachyon $T$ of the Type IIB system is a complex field and the minimum 
of its potential does not consist any more of two separate points ($\pm
T_0$), but lies along a circle of radius $|T_0|$.
Thus a kink configuration can not be stable 
because there is not any topological constraint 
that forbids its decay
into the trivial solutions $T=T_0$ or $T=-T_0$. 
The possibility of this decay is
precisely
described by the tachyonic mode (\ref{ntac2}) that is present in the Type
IIB theory, but not in the Type I model where it is projected out by $\Omega$.

In conclusion we have shown that a D1/anti-D1 pair with a $Z_2$ Wilson
line $\theta=\pi$ on the anti-brane and with a ``tachyonic'' Wilson line $A=1$
at the critical radius $R_c$ describes a D-particle which is non-BPS
and unstable in Type IIB theory. However, by performing the
$\Omega$ projection the tachyon instability is removed and the 
non-BPS D-particle of Type I is a stable configuration.

\vskip 1.5cm
\sect{The Type I D-particle in Minkowski space}
\label{dec}
\vskip 0.5cm
%%%%%%%%%%%%%%%%%%%%%
% Section 6
%%%%%%%%%%%%%%%%%%%%%

From the qualitative considerations of \secn{d0}, it emerged that a
the D1/anti-D1 pair with a $Z_2$ gauge Wilson line has the 
right quantum numbers to describe the $SO(32)$ spinor states which are 
present at the first massive level of the perturbative heterotic
spectrum. In the 
previous section we learned that at the critical radius 
$R_c=\sqrt{\alpha'/2}$ it is possible
to describe exactly the tachyonic background living on the system and to
study the conformal field theory of the open strings ending on it.
Now it is crucial to show that, once the tachyon has condensed,
the stability and the properties of the resulting bound state do not 
depend on the value of the radius and that it is possible to smoothly
take the decompactification limit $R\to \infty$ to describe a
D-particle of Type I in Minkowski space. To show this, we will proceed in 
the framework
of perturbation theory: within the conformal field theory at critical
radius $R=R_c$, we describe a small variation of $R$ by
inserting in the various amplitudes
the operator ${\cal V}_{R}$
associated to the radius deformation. 
For instance, by calculating the one
point function of the vertex operator (\ref{vat0}), one can check the
stability of our vacuum configuration; in fact, if the background
satisfies the classical equations of motion all tadpoles are vanishing.
The same amplitude with a ${\cal V}_{R}$ insertion describes the first
order correction to the stability of
this background arising from the $R_c\to R_c+\delta R$ variation. We 
have claimed that the value $A=1$ is special, since it should represent
a stable configuration also beyond the critical radius; if this is the 
case, it is clear that the tadpoles cancellation must not be effected 
by the variation of $R$ and thus the amplitude
\begin{equation}
\label{tadt}
T_\phi = \langle {\rm T} \Big( {\cal V}_{\cal T}(x)\,{\cal 
V}_{R}(z)\Big) \rangle_{A}
\end{equation}
should vanish for $A=1$ \footnote{Because of projective invariance, the 
correlation function in \eq{tadt} does not depend on the positions of 
the vertex operators and it is convenient to exploit this fact by 
fixing $x=1$ and $z=\ii$.}.

The explicit form of the marginal operator ${\cal V}_R$ is related to 
the graviton vertex at zero-momentum which, for instance, in the bosonic 
theory is given by
\begin{equation}
\label{rd0}
{\cal V}_R = \partial X(z)\; \bar\partial\widetilde{X}(\bar z)~.
\end{equation}
The same result holds also for superstrings when ${\cal V}_R$ is 
written in the $(0,0)$ - superghost picture (that is, when it does not 
carry either left or right superghost charge), while, in the
$(-1,-1)$ - picture, the bosonic 
coordinate $X$ is substituted by $\psi$ and the vertex reads
\begin{equation}
\label{rd-1}
{\cal V}_R^{(-1,-1)} = \psi(z) \; \widetilde{\psi}(\bar z)~.
\end{equation}
A second ingredient which is necessary for the computation of 
the amplitude
(\ref{tadt}) is the definition of the expectation value $\langle \ldots
\rangle_A$ in presence of the non trivial ``tachyonic'' background.
From the analysis of the previous section we already know that in terms 
of the $(\eta,\phi)$ variables the 
``tachyonic'' background can be easily taken into account by shifting 
the $\phi$-momentum of charged states according to \eq{shift}.
However, this is {\em not} the only
modification to disk amplitudes (like the 
one in \eq{tadt}) induced by the presence of $A$ because there is
one more subtlety. In 
fact, one has
to remember that in general, when closed string vertex operators are 
present, a
gauge field $Y$ couples also to the total winding number carried by the
closed external states \cite{g91}. Thus, in the presence of a $Y$
background, the 
expectation value on a disk reads as follows
\begin{equation}
\label{exY}
\langle \ldots \rangle_Y =  \langle\, \ldots\, \ex^{{\rm i} \,Y_I \oint
d\sigma^\a\, \partial_\a X^I} \rangle~~.
\end{equation}
Note that this modification is necessary to ensure the invariance of the
disk amplitude under a T-duality that transforms the constant gauge 
field $Y$ in the position of the D-brane. In our case, when the $\cal T$
background is written in the $(\eta,\phi)$ language, it has exactly the
form of a constant gauge field and so, as \eq{wilsa} suggests,
the expectation value
$\langle\ldots\rangle_A$ should be defined as in \eq{exY}, namely
\begin{equation}
\label{exA}
\langle \ldots \rangle_A =  \langle\, \ldots\,{\rm Tr}\left((\ldots)\;
\ex^{{\rm i} {A\over 2\sqrt{2\a'}}\oint d\sigma \, 
\partial_\sigma \phi \,\otimes\,\sigma^1}\right) \rangle~,
\end{equation}
where the Chan-Paton factors of the external open strings have to be 
inserted inside the trace. A final observation should be made about the 
time ordering T in \eq{tadt}: in order to avoid ambiguities among the 
open and closed vertex operators it is useful to perform the change of
variables \cite{ademetal74}
\beq \label{z'}
z \rightarrow z'=-{z-{\rm i} \over z+{\rm i}}~~~.
\eeq
This transformation maps the upper half $z$-plane into the circle of
unit radius and the lower part of the $z$-plane outside the same circle.
Thus the closed string vertices are split and all the holomorphic
parts are placed on the right, while all the antiholomorphic ones are 
placed on the left; the open string vertices are inserted in between,
since they are forced to stay exactly on the unit circle.

Now we are in the position of evaluating the correlation 
function in \eq{tadt}; the first step is to translate the
operator (\ref{rd-1}) in the $(\phi,\eta)$ language in order to be able 
to use the definition (\ref{exA}) for the expectation value in presence 
of an
non trivial $A$. We have already performed this transformation at level
of states (see \eq{4ns}). Then, by repeating the same steps and 
exploiting the freedom to fix the punctures at $0$, $1$ and $\infty$, it 
is easy to see that \eq{tadt} can be explicitly rewritten as
\begin{equation}
\label{tadt2}
T_\phi \sim \langle \Big(\ex^{{\ii \over\sqrt{2\a'}}\widetilde\phi} - 
\ex^{-{\ii \over\sqrt{2\a'}}\widetilde\phi} \Big)\Big|_\infty 
\partial\phi\Big|_1 \Big(\ex^{{\ii \over\sqrt{2\a'}}\phi} - \ex^{-{\ii 
\over\sqrt{2\a'}}\phi} \Big)\Big|_0 {\rm Tr} \left(\sigma^1 \ex^{{\rm 
i} {A\over 2\sqrt{2\a'}}\oint d\sigma \,
\partial_\sigma \phi \,\otimes\,\sigma^1}\right) \rangle~,
\end{equation}
where the left and the right fields $\phi$ and $\widetilde \phi$
must be identified like in any disk amplitude.
The structure of this equation is identical to that of a three-point
function among open strings. So it is easy to see that two of the four
terms arising from \eq{tadt2} are trivially zero because of
$\phi$-momentum conservation; in fact, as stressed in Appendix A, 
the new boson $\phi$ defined in \eq{bos2} satisfies Neumann boundary 
conditions so that its momentum can not flow away from the disk 
boundary. The only two terms that give a non zero result are 
\begin{equation}
-\Big( \ex^{{\ii \over\sqrt{2\a'}} \widetilde
\phi}\Big|_\infty \partial\phi\Big|_1 
\ex^{-{\ii\over\sqrt{2\a'}}\phi}\Big|_0 \Big) 
\label{2ter}
- \Big( \ex^{-{\ii \over\sqrt{2\a'}} \widetilde\phi}\Big|_\infty
\partial\phi\Big|_1 \ex^{{\ii \over\sqrt{2\a'}}\phi}\Big)\Big|_0\Big)~~,
\end{equation}
so that \eq{tadt2} reduces to
\begin{eqnarray}
\label{tadt21}
T_\phi & \sim & -~\langle ~\ex^{{\ii \over\sqrt{2\a'}}\widetilde\phi}
\Big|_\infty  ~          
\partial\phi\Big|_1 ~\ex^{-{\ii  
\over\sqrt{2\a'}}\phi}
\Big|_0 ~{\rm Tr} \left(\sigma^1 \ex^{{\rm   
i} {A\over 2\sqrt{2\a'}}\oint d\sigma \,                                
\partial_\sigma \phi \,\otimes\,\sigma^1}\right) \rangle            
\nonumber \\
&-&\langle ~\ex^{-{\ii \over\sqrt{2\a'}}\widetilde\phi}   
\Big|_\infty  ~                                                   
\partial\phi\Big|_1 ~\ex^{{\ii                                   
\over\sqrt{2\a'}}\phi}                                            
\Big|_0 ~{\rm Tr} \left(\sigma^1 \ex^{{\rm                        
i} {A\over 2\sqrt{2\a'}}\oint d\sigma \,                          
\partial_\sigma \phi \,\otimes\,\sigma^1}\right) \rangle ~.
\end{eqnarray}
To proceed further, we first notice that 
$\partial\phi$ basically reads the momentum carried by the second
exponential factor inside the correlators of \eq{tadt21}; hence $T_\phi$ 
becomes 
\begin{equation}
\label{tadt3}
T_\phi \sim \langle 
\left[\ex^{-{\ii\over\sqrt{2\a'}}(q - \widetilde{q})} -
\ex^{{\ii\over\sqrt{2\a'}}(q - \widetilde{q})}\right]
{\rm Tr} \left( \sigma^1 \ex^{{\rm i}
{A\over 2\sqrt{2\a'}}\oint d\sigma \,
\partial_\sigma \phi \,\otimes\,\sigma^1}\right)\, \rangle~,
\end{equation}
where $q$ and $\widetilde q$ are the zero modes of $\phi$ and 
$\widetilde \phi$.
Let us recall that in our conventions~\footnote{Our conventions on
closed                 
string fields are slightly different from the one of 
Ref.~\cite{polB}; in      
particular the left and the right components of the coordinate $X^\mu$    
are defined as $X^\mu(z,\bar z) = \shalf[X^\mu(z) + X^\mu(\bar z)]$,      
where $z=\ex^{2(\tau+\ii\sigma)}$. This implies that $k_L$ and $k_R$      
momenta are quantized as follows                                          
\[
{k_L^\mu} = {1\over 2}\left({n^\mu\over R}                                
+ {w^\mu R\over \a'}\right)~~,~~~                                         
{k_R^\mu}  =  {1\over 2}\left({n^\mu\over R}                              
- {w^\mu R\over \a'}\right)~.                                             
\]}, 
a vertex operator
describing a closed string state with Kaluza-Klein index $n_\phi$ and 
winding number $w_\phi$, contains the term 
\begin{equation}
\label{w21}  
\ex^{\ii k_L \phi}~\ex^{k_R\widetilde\phi} \sim
\ex^{{\ii\over\sqrt{2\a'}}(n_\phi+\frac{w_\phi}{2})q_\phi}
~\ex^{{\ii\over\sqrt{2\a'}}(n_\phi-\frac{w_\phi}{2}){\widetilde q}_\phi}
\end{equation}                                                    
According to this formula, we easily see that 
the two terms in the square brackets of \eq{tadt3} correspond
respectively to $(n_\phi=0,w_\phi=2)$ and $(n_\phi=0,w_\phi=-2)$.
Therefore, even if the left-right identification induced by the disk
topology of the amplitude implies that both terms are 
simply equal to one and give no
direct contribution, their presence is not trivial. In fact, since 
they carry different winding numbers, they give a different contribution
to the contour integral $\oint \partial\phi$
because
\begin{equation}
\label{contint}
\frac{A}{2\sqrt{2\alpha'}}\oint d\sigma \partial_\sigma \phi
=\pi\frac{w_\phi}{2}A
\end{equation}
At this point the evaluation of \eq{tadt3} simply reduces to that of the
trace factor and one gets
\begin{equation}
\label{tadt4}
T_\phi \sim \Bigg[-{\rm Tr}\Big(\sigma^1 \ex^{\ii A\pi\sigma^1 }\Big) +
{\rm Tr}\Big(\sigma^1 \ex^{-\ii A\pi\sigma^1 }\Big)\Bigg] \sim
\sin{(\pi A)}~;
\end{equation}
this result is zero for $A=1$ showing that, for this particular 
background, the ``tachyon'' tadpole is still vanishing, even if the 
radius has been deformed away from the critical value.

This same approach can be used to study how the mass$^2$ of the 
``tachyonic'' state (\ref{calt}) is modified by a radius deformation; in
this case, the relevant amplitude is the two-point function in presence 
of an operator ${\cal V}_R$, namely
\begin{equation}
\label{mr0}
m_\phi =
\langle {\rm T} \Big({\cal V}_{\cal T}(x)\,{\cal V}_{\cal T}(y)\,  
{\cal V}_{R}(z)\Big) \rangle_{A} 
\end{equation}
Exploiting the projective invariance to fix $y=1$ and $z=\ii$, 
performing the change of variable of \eq{z'} and using the form
of the vertex operators in terms of $\phi$ and $\widetilde\phi$, we obtain
\begin{equation}
\label{mr}
m_\phi 
\sim \langle \Big(\ex^{{\ii
\over\sqrt{2\a'}}\widetilde\phi}
- \ex^{-{\ii \over\sqrt{2\a'}}\widetilde\phi} \Big)\Big|_\infty 
\partial\phi\Big|_1\partial\phi\Big|_x \Big(\ex^{{\ii 
\over\sqrt{2\a'}}\phi} - \ex^{-{\ii \over\sqrt{2\a'}}\phi} \Big)\Big|_0 
{\rm Tr} \left( \ex^{{\rm i} {A\over 2\sqrt{2\a'}}\oint 
d\sigma \,\partial_\sigma \phi \,\otimes\,\sigma^1}\right) 
\rangle ~,
\end{equation}
where the Chan-Paton factors of the two open strings cancel out, since
$(\sigma^1)^2 =1$. As before, some terms in this equation are killed by
$\phi$-momentum conservation and one is left with only two terms 
\begin{equation}\label{mr2} 
m_\phi \sim \langle \Bigg(
\Big[\ex^{-{\ii\over\sqrt{2\a'}}(q_\phi - \widetilde{q}_\phi)} \Big] +
\Big[\ex^{{\ii\over\sqrt{2\a'}}(q_\phi - \widetilde{q}_\phi)} \Big] 
\Bigg) {\rm Tr} \left( \ex^{{\rm i}{A\over 2\sqrt{2\a'}}\oint d\sigma\,
\partial_\sigma \phi \,\otimes\,\sigma^1}\right) \rangle~.
\end{equation}
Note that the sign between the two exponential factors is different from 
that of \eq{tadt3}, because now we have two $\partial\phi$ insertions.
Evaluating the trace over the Chan-Paton factors, one finally obtains
\begin{equation}
\label{mr3}
m_\phi \sim \cos{(\pi A)}~.
\end{equation}
This simple result contains a crucial information: the mass behavior of 
our ``tachyonic'' state is very different if one deforms the
compactification radius around the trivial configuration $A=0$ or around 
the kink solution $A=1$; indeed
\begin{equation}
\label{opp}
\langle {\rm T} \Big({\cal V}_{\cal T}(x)\,{\cal V}_{\cal T}(y)\, 
{\cal V}_{R}(z)\Big) \rangle_{A=0} = -
\langle {\rm T} \Big( {\cal V}_{\cal T}(x)\,{\cal V}_{\cal T}(y)\, 
{\cal V}_{R}(z)\Big) \rangle_{A=1}~.
\end{equation}
We already know that in the absence of a ${\cal T}$ background, the 
D1/anti-D1 pair develops a tachyonic instability when we take the 
decompactification limit $R\to\infty$; this fact can now be seen 
directly using the result we have just obtained. Indeed, from
\eq{masslesstach}, it is easy to realize that the mass$^2$ of the two 
lightest modes ($n=0$, $n=-1$) considered in the previous section, 
{\em decreases} when the radius $R$ is increased and becomes negative 
above the critical value $R_c$. On the other hand, \eq{opp} tells us 
that this behavior is reversed when $A=1$ and that the system is stable
if $R\geq R_c$. Thus, for $A=1$ one can take the decompactification
limit without encountering any tachyonic instability.

Even if here we have considered only the effects of a first order 
variation of the radius (that is only one ${\cal V}_{R}$ was inserted),
the results we have obtained are quite general and a 
similar pattern is found
also when several ${\cal V}_{R}$ insertions are taken into account. In 
fact, as is clear from the two explicit examples just described, the 
relevant part in the disk amplitudes describing the effect of a radius 
deformation comes from the factor
\begin{equation}
\label{trf}
\ex^{{\rm i} {A\over 2\sqrt{2\a'}}\oint d\sigma \, \partial_\sigma \phi
\,\otimes\,\sigma^1} = \cos{\left(\pi {w_\phi\over 2} A \right)} +
\sigma^1 \sin{\left(\pi {w_\phi\over 2} A\right)}~,
\end{equation}
which has to be inserted in the correlation functions according to
\eq{exA}. In
particular, ${\cal V}_{R}$ can only give rise to
combination with even winding number (just like in \eq{mr2}), so that 
the $\sin(\ldots)$ term always vanishes with our preferred value $A=1$.
On the other hand, the $\cos(\ldots)$ term simply changes the sign in
front of operators with odd  $w_\phi/2$. 
It is very instructive to use
this input directly in the vertex describing the marginal deformation of
the radius. To this aim, we can focus on \eq{rd0}, since, after the 
insertion of the
first ${\cal V}_{R}$ in the $(-1,-1)$ picture, the superghost anomaly of
the disk is compensated and for all other insertions one should use
vertex  ${\cal V}_{R}$ in the $(0,0)$. In the $(\phi,\eta)$
language this vertex reads
\begin{eqnarray}
\label{rdoep}
{\cal V}_R & = & \partial X(z)\; \bar\partial\widetilde{X}(\bar z) =
-(2\a') \xi\eta\widetilde{\xi}\widetilde{\eta} \\ \nonumber
& = & \sqrt{2}\a' \eta\widetilde{\eta}
\Big[\ex^{{\ii\over\sqrt{2\a'}}\phi} + 
\ex^{-{\ii\over\sqrt{2\a'}}\phi} \Big] 
\Big[\ex^{{\ii\over\sqrt{2\a'}}\widetilde\phi} +
\ex^{-{\ii\over\sqrt{2\a'}}\widetilde\phi} \Big]
\\ \nonumber
& = & \sqrt{2}\a' \eta\widetilde{\eta}\left(
\ex^{{\ii\over\sqrt{2\a'}}(\phi - \widetilde{\phi})} +
\ex^{-{\ii\over\sqrt{2\a'}}(\phi - \widetilde{\phi})} +
\ex^{{\ii\over\sqrt{2\a'}}(\phi + \widetilde{\phi})} +
\ex^{-{\ii\over\sqrt{2\a'}}(\phi + \widetilde{\phi})} \right)~.
\end{eqnarray}
Since the Wilson line $A=1$ simply changes the sign in front of the 
exponentials with $\pm(\phi - \widetilde{\phi})$, we can forget the presence 
of the non-trivial background in the trace factor, provided that we use 
the following modified vertex operator
${\cal V}_{R}'$ 
\begin{eqnarray}
\label{mvr}
{\cal V}_{R}' & = &  \sqrt{2}\a' \eta\widetilde{\eta}\left(
\ex^{{\ii\over\sqrt{2\a'}}(\phi - \widetilde{\phi})} -
\ex^{-{\ii\over\sqrt{2\a'}}(\phi - \widetilde{\phi})} +
\ex^{{\ii\over\sqrt{2\a'}}(\phi + \widetilde{\phi})} -
\ex^{-{\ii\over\sqrt{2\a'}}(\phi + \widetilde{\phi})} \right) \nonumber \\
& = &\sqrt{2}\a' \eta\widetilde{\eta}
\Big[\ex^{{\ii\over\sqrt{2\a'}}\phi} - 
\ex^{-{\ii\over\sqrt{2\a'}}\phi} \Big] 
\Big[\ex^{{\ii\over\sqrt{2\a'}}\widetilde\phi} - 
\ex^{-{\ii\over\sqrt{2\a'}}\widetilde\phi} \Big]~.
\end{eqnarray}
If one uses as independent variables the fields $\phi'$ and $\xi$ 
given in \eq{bos3}, this expression takes a very 
simple form 
\begin{equation}
\label{mvr'}
{\cal V}_{R}' = - \Big(\ii \sqrt{2\a'}\;\eta\,\psi\Big)
\Big(\ii \sqrt{2\a'}\;\widetilde\eta \, \widetilde\psi\Big) =
-\partial\phi'\,\bar\partial\widetilde\phi'~.
\end{equation}
Thus, the presence of the ``tachyon'' vacuum expectation value $A=1$ 
effectively substitutes $X^1$ with $\phi'$ in the definition of the 
marginal operator related to radius deformations and
transforms it into $-\partial\phi'\,\bar\partial\widetilde\phi'$. The 
minus sign in this expression has a crucial importance; in fact, this 
means that increasing
the radius of the $X^1$ coordinate ($R_c\to R_c +\delta R$) is 
equivalent, at $A=1$, to {\em decrease} the radius of the $\phi'$ coordinate 
($R_c\to R_c - \delta R$), so that the decompactification limit can be 
also seen as a $R_{\phi'}\to 0$ limit. It is by now well known that the 
behavior of open strings compactified on a small circle is more easily 
understood by making a T-duality transformation. In our case, the field 
$\phi'$ satisfies Neumann boundary conditions like $\phi$, since both
come from the bosonization procedure described in Appendix A; therefore 
after a T-duality, $\phi'$ becomes a Dirichlet coordinate $\phi'_D$
compactified on a circle of radius $R_{\phi'_D}=\alpha'/R_\phi$ and 
the decompactification limit of the original space-time translates into
the $R_{\phi'_D} \to \infty$ limit. Thus in Minkowski
space the bound state between a D1 brane and its anti-brane
is described by one Neumann field ($X^0$) and by nine Dirichlet
coordinates $(\phi'_D, X^2,
\ldots X^9)$. This is an indication that, even if the starting point of 
our analysis was a system with only a $SO(8)$ symmetry manifest, the 
stable configuration in the flat space has an enlarged $SO(9)$ symmetry;
in fact, even if we do not show it explicitly here, it is not difficult 
to see that $\phi'$ stays on the same footing of other Dirichlet 
coordinates:
for instance, the full spectrum of open strings living on the bound 
state is invariant under the exchange $(\phi'_D,\xi)\leftrightarrow 
(X^2,\psi^2)$. The $SO(9)$ symmetry of the final configuration in the 
decompactified space corresponds to the little group of massive particle
living in ten dimensions, so that the resulting configuration can be 
interpreted as a D-particle in Type I theory. 

Since now we know the effect of radius deformation on the conformal 
theory living on the D1/anti-D1 system after tachyon condensation, 
we can return 
to the problem of the mass $M$ of the new D-particle state and show 
that for $A=1$ it does not depend on the radius. This can be shown by 
means of the same approach used by J. Polchinski in Ref.~\cite{polc95} 
to find the
mass of the D-branes in Type II theories. One starts by computing 
the vacuum energy between two identical systems and then extracts the 
part corresponding to the exchange of closed strings in the NS-NS spin 
structure. In the limit of very large separation, only the massless 
states contribute and this string result, which depends only on
$\a'$ and $g$, has to be equal to the field theory answer which instead
is given in terms of the mass $M$ and of the Newton's constant $G$. 
Using this method, we now check that, in the 
absence of a ${\cal T}$ background ($A=0$), the interaction between two 
D1/anti-D1 pairs give the expected answer. 
Newton's law (in ten dimension) tells that the potential energy
between two masses $M$ placed at a distance $\ell$ from each other
scales as 
\begin{equation}
\label{newlaw}
U~\sim ~G\, {M^2\over \ell^6}~,
\end{equation}
where the Newton's constant is proportional to $1/R$ since one dimension 
($X^1$) is compactified. Also the
string result scales as $1/\ell^6$; in fact, the vacuum energy between 
two $p$-extended objects at a distance $\ell$ from each other is
\begin{equation}
\label{ustring}
U_{\rm string}^{A=0} ~\sim~ V_{p+1}\,\frac{1}{\ell^{7-p}}
\end{equation}
where $V_{p+1}$ is the world-volume of the $p$-brane. In our case
$p=1$ and $V_2$ scales as $R$.
Thus, by equating the $R$ dependence of Eqs. (\ref{newlaw}) and  
(\ref{ustring}) for $p=1$, we find, as
expected, that the mass of the system is 
proportional to the radius: $M\sim R$. In the presence of the kink 
solution ($A=1$), the gravitational interaction is 
still described
by \eq{newlaw}, but the string computation is different. 
In fact, from 
\eq{mvr'} we know that the effective radius $R_\phi'$ one should 
use in the
conformal field theory analysis is the inverse of the physical radius 
$R$ along
the $X^1$ direction. Thus, for $p=1$ we get
\begin{equation}
\label{ustring1}
U_{\rm string}^{A=1}~\sim~V'_{2}\,\frac{1}{\ell^6}~\sim~
\frac{1}{R}\,\frac{1}{\ell^6}
\end{equation}
Equating the $R$ behavior of Eqs. (\ref{newlaw}) and (\ref{ustring1})
we now find that at $A=1$ the mass $M$ of the D1/ant-D1 pair is {\em 
independent} of $R$. Thus, the mass of the Type I D-particle
described by this system in the flat Minkowski space is the same as the 
one computed at the critical radius, {\em i.e.}
\begin{equation}
\label{massfin}
M=\frac{1}{\sqrt{\alpha'}\,g_I}
\end{equation}
as we have seen in \eq{ormass}.

In these notes we have mainly concentrated on the construction
of the stable non-BPS D-particle of Type I which is predicted and
required by the heterotic/Type I duality. However, the techniques
we have used and the strategy we have discussed
can be generalized and applied to other systems and theories.
Even if the entire construction seems quite complicated and 
technically involved,
the emerging picture is very satisfactory and appealing
and makes manifest the fact that the presence of a tachyonic
field in a string theory is not necessarily a sign of inconsistency. 
Of course there are many issues and developments that have not
been covered in these notes but which are very interesting, such as for example the
construction of other stable non-BPS D-branes of Type I 
\cite{WITTEN,Gallot} or
the relation between the D-brane charge and K-theory \cite{WITTEN,HORAVAK}.
Moreover, there are still some open problems which remain to be solved:
among them we can mention the problem of understanding
the mechanism of tachyon condensation
directly at the level of the effective theory of the branes which
in turn requires also
a better understanding of the fate of the various gauge fields living on
D-brane world volume \cite{WITTEN,Yi}. A progress in this direction 
presumably
could help to understand the mechanism of tachyon condensation also 
in those cases where the tachyon appears in the closed string spectrum.

%%%%%%%
\vskip 2cm
{\large {\bf {Acknowledgements}}}
\vskip 0.5cm
It is a pleasure to thank P. Di Vecchia, M. Frau, L. Gallot, I. 
Pesando. S. Sciuto and P. Strigazzi for many useful discussions,
and the organizers and the participants of
the School {\it Quantum Aspects
of Gauge Theories, Supersymmetry and Unification} in Leuven
(Belgium) January 1999 for their stimulating
questions and observations. R. Russo wishes also
to thank A. Bilal and D. Matalliotakis for 
many valuable suggestions and observations.

%%%%%%%

\vskip 1.5cm
\appendix{\Large {\bf {Appendix A}}}
\vskip 0.5cm
\renewcommand{\theequation}{A.\arabic{equation}}
\setcounter{equation}{0}
\noindent
%%%%%%%%%%%%%%%%%%%%%%%%%
% insert APPENDIX A
%%%%%%%%%%%%%%%%%%%%%%%%%
In Section \ref{cr} we have extensively used the well known equivalence 
between a compact bosonic field and two real fermions.
This equivalence is exploited in string theory in many contexts,
but, even if the basic idea is always encoded by \eq{bos}, there are 
some differences among various cases. 
For instance, the bosonization procedure relates the two formulations 
of heterotic theory mentioned in Section \ref{pert}; however, in that 
case, the bosonic fields satisfied periodic boundary conditions and
the critical radius was $\sqrt{\a'}$. On the other hand, if Neumann 
boundary conditions are imposed, the equivalence holds for 
$R=\sqrt{2\a'}$. In fact, by looking at \eq{bos}, it is clear that the 
first fermionic state one can construct by acting with $(\xi\pm\ii
\eta)$ on the vacuum corresponds, in the bosonic description, to a 
scalar state of momentum $\pm 1/\sqrt{2\a'}$. Since higher fermionic
states are obtained by introducing other factors of $(\xi\pm\ii \eta)$,
in the bosonic language one has to consider states whose momentum $k$
is $n/\sqrt{2\a'}$ with $n\in\interi$. Further evidence that the two
formulations really describe the same theory is provided by the
computation of the partition function. In terms of the field $X^1$, the 
partition function factorizes in a sum over the allowed momenta and 
in the oscillator contribution
\begin{equation}
\label{pfbos}
{\rm Tr}\left(q^{2 L_0}\right) = \sum q^{2\a' k^2} \prod_{n=1}^\infty 
{1\over (1-q^{2n})}~.
\end{equation}
However, because of the particular form of the momentum, the sum 
take a very simple form and can be rewritten as a product
\begin{equation}
\label{m1}
\sum_{n\in\interi} q^{n^2} = \prod_{n=1}^\infty
\left(1-q^{2n}\right) \left(1+q^{2n-1}\right)^2~,
\end{equation}
and so \eq{pfbos} can be recast in a different form
\begin{equation}
\label{pff}
%{\rm Tr}\left(q^{2 L_0}\right) = 
\sum q^{2\a' k^2} \prod_{n=1}^\infty 
{1\over (1-q^{2n})} = \prod_{n=1}^\infty
\left(1+q^{2n -1}\right)^2~,
\end{equation}
that directly matches the partition function of two real NS fermions.
However, this is not exactly the result we needed in Section \ref{cr}, 
where the bosonic field was compactified on a circle of radius 
$R_c=\sqrt{\a'/2}$ and could have integer or half-integer values of the 
momentum according to various Chan-Paton factors (but, of course not 
both together). If one looks at the states proportional to $\sigma^1$
or $\sigma^2$, it is easy to realize that they simply are a subset of 
those that can be constructed when $R=\sqrt{2\a'}$; in particular their 
momentum is of the form $(2n+1)/\sqrt{2\a'}$. This consideration 
suggests that, at the critical radius we are interested in, the only
bosonic states that can be constructed correspond to fermionic 
states with an {\em odd} number of $\xi$ and $\eta$ oscillators.
On the other hand, the states proportional to $\sigma^3$ or$\one$ are 
periodic: this implies that their momentum is integer in terms of
$R_c$ and thus it can take only even values in terms of $R=\sqrt{2\a'}$. 
In this case the corresponding fermionic states contain an {\em even} 
number of $\xi$ and $\eta$ oscillators.
Thus it turns out that the states with half-integer momentum
of the bosonic field correspond to states with the usual GSO projection
in the fermionic language; on the contrary, when the momentum is
quantized in integer multiples of $1/R_c$, one reproduces the sector
with the "wrong" GSO projection of the two real fermions. 
This can be checked again at the level of partition function. Let us 
focus, for example, on the $\sigma^1$ $\sigma^2$ sector. Here the sum
over the allowed momenta in \eq{pfbos} takes a slightly different form, 
but nevertheless it can be transformed in the expected partition 
function by using again the identity (\ref{m1})
\begin{eqnarray}
\sum_{n\in \interi} q^{(2n+1)^2} & =& \frac{1}{2} 
\sum_{n\in\interi} \left( q^{n^2} - (-q)^{n^2}\right) 
\\ \nonumber & = &
\frac{1}{2} \prod_{n=1}^\infty \left(1-q^{2n}\right)
\left[\left(1+q^{2n-1}\right)^2 - \left(1-q^{2n-1}\right)^2
\right]~.
\end{eqnarray}
It is possible to resume this result in a simple requirement, which also 
allows us to treat the states with different Chan-Paton factors in the 
same way. In fact, we can think that the momentum of the bosonic field 
$X^1$ is always allowed to assume both integer and half-integer values, 
but, then, we introduce a new discrete symmetry $h$ that selects 
physical states. From previous analysis, it is clear that $h$ acts on
$\xi$ and $\eta$ as a GSO projection; then from
Eqs.~(\ref{bos2})-(\ref{bos3}) one can derive how $h$ translates in 
terms of the rebosonized fields. For instance, by
taking $\phi$ and $\eta$ as independent fields, it is easy to 
see from \eq{bos2} that $h$ transforms $:\ex^{{\ii\over \sqrt{2\a'}}
\phi}:$ in $(-:\ex^{-{\ii\over \sqrt{2\a'}} \phi}:)$. The explicit form of
$h$ in the different languages is reported in \tbl{hproj}.
\begin{table}[htp]
$$
\begin{array}{|c|} 
\hline h \\ \hline \\
\left\{\begin{array}{l}
X \to X + 2\pi\sqrt{\a'\over 2} \\
\psi^\mu \to \psi^\mu \\
\ket{0}_{-1} \to \ket{0}_{-1}  \\
\lambda \to \sigma^3\, \lambda\,\sigma^3 \\
\end{array}
\right\} \Leftrightarrow
\left\{\begin{array}{l}
\xi \to -\xi %\;,~
\\ \eta \to -\eta \\
\psi^\mu \to \psi^\mu \\
\ket{0}_{-1} \to \ket{0}_{-1} \\
\lambda \to \sigma^3\, \lambda\,\sigma^3 \\
\end{array}
\right\} \Leftrightarrow
\left\{\begin{array}{l}
\phi \to -\phi + 2\pi \sqrt{\a'\over 2} \\
\eta \to -\eta %\;,~%
\\ \psi^{\mu\not= 1}\to \psi^{\mu\not= 1} \\
\ket{0}_{-1} \to \ket{0}_{-1} \\
\lambda \to \sigma^3\, \lambda\,\sigma^3 \\
\end{array}
\right\}
\\ \hline
\end{array}
$$
\caption{The discrete symmetry $h$ that defines the physical states 
according to the different Chan-Paton factors: physical states are 
invariant under these transformations. \label{hproj}} 
\end{table}

\noindent 
In a similar way one can also write down the GSO projection in terms
of sets of independent fields and derive the results of \tbl{vargso}.
\begin{table}[htp]
$$
\begin{array}{|c|} 
\hline (-1)^F \\ \hline \\
\left\{\begin{array}{l}
X \to X \\
\psi^\mu \to -\psi^\mu \\
\ket{0}_{-1} \to -\ket{0}_{-1}  \\
\lambda \to \sigma^3\, \lambda\,\sigma^3 \\
\end{array}
\right\} \Leftrightarrow
\left\{\begin{array}{l}
\xi \to \xi % \;,~
\\ \eta \to \eta \\
\psi^\mu\to -\psi^\mu \\
\ket{0}_{-1} \to -\ket{0}_{-1} \\
\lambda \to \sigma^3\, \lambda\,\sigma^3 \\
\end{array}
\right\} \Leftrightarrow
\left\{\begin{array}{l}
\phi \to -\phi \\
\eta \to \eta \\
\psi^{\mu\not= 1}\to - \psi^{\mu\not= 1} \\
\ket{0}_{-1} \to -\ket{0}_{-1} \\
\lambda \to \sigma^3\, \lambda\,\sigma^3 \\
\end{array}
\right\} \\ \hline
\end{array}
$$
\caption{The generalized GSO projection defined in 
Eq.~(\ref{gsocp}) 
written in terms of different fields: physical states
are invariant under these transformations.
\label{vargso}}
\end{table}

Of course, the boundary conditions of the new bosonic field $\phi$ 
depends on those of $\xi$ and $\psi$. For instance, if both fermions
are in the NS sector, it is sufficient to read \eq{pff} backwards to see
that $\phi$ is an usual Neumann field. However, while $\xi$ is always a 
NS field, $\psi$ can also be in the R sector: in this case the
combination of the two gives a bosonic field with Dirichlet boundary 
conditions on one endpoint and Neumann boundary conditions on the other.
Thus, $\phi$ is half-integer moded and can not carry momentum,
thus its partition function is simply $(\prod (1-q^{2m-1}))^{-1}$. 
The partition function in the fermionic description can be recast in 
this form showing again that the two representations are equivalent
\begin{equation}
\label{nsr} % Z_R & = &
Z_R = \prod_{m=1}^\infty \left(1+q^{2m-1} \right) 
\left(1+q^{2m}\right) % \\ \nonumber & = & 
= \prod_{n=1}^\infty \left(1+q^n \right) = 
\prod_{m=1}^\infty {1\over \left(1-q^{2m-1} \right)} ~.
\end{equation}
Finally a useful observation is that the two discrete symmetries 
$(-1)^F$ and $h$ can be combined so that the resulting transformation is
much simpler in the $(\phi$, $\eta$) description.
\begin{table}[htp]
$$ %\left.
\begin{array}{|c|}
\hline (-1)^F\,h \\  \hline \\
\phi \to \phi + 2\pi \sqrt{\a'\over 2} \\
\eta \to -\eta \\
\psi^{\mu\not= 1}\to - \psi^{\mu\not= 1} \\
\ket{0}_{-1} \to -\ket{0}_{-1} \\
\lambda \to \lambda \\ \hline
\end{array}
%\right\} (-1)^F\,h
$$
\caption{\label{Fh}The $(-1)^F\,h$ in terms of $\phi$ and $\eta$}
\end{table}

\noindent
Since the physical states have to be even under both $(-1)^F$ and $h$ 
separately, they are also even under $(-1)^F\,h$ and either $(-1)^F$ or 
$h$, and indeed the two requirements are equivalent.

\newpage

%\bibliographystyle{myu}
%\bibliography{bibmam}
%
%\end{document}

\end{document}